\documentclass[final,3p,times]{elsarticle}

\usepackage[utf8]{inputenc}

\title{An overview of experimental results from ultra-relativistic heavy-ion collisions at the CERN LHC: bulk properties and dynamical evolution}

\usepackage{graphicx}
\usepackage{amssymb}
\usepackage{amsmath}
\usepackage{bm}
\usepackage{color}
\usepackage{float}
\usepackage{caption}
\biboptions{sort&compress}
\usepackage{xspace}
\usepackage{tabularx}
\usepackage{multirow}
\newcommand{\Pt}{$p_{\rm{T}}$}
\newcommand{\pt}{$p_{\rm{T}}$}
\newcommand{\pT}{$p_{\rm{T}}$}

\newcommand{\detadphi}{$\Delta\eta\Delta\varphi$}

\newcommand{\jpsi}{J/$\psi$}

\newcommand{\pp}{pp}

\newcommand{\pPb}{\mbox{p--Pb}}
\newcommand{\PbPb}{\mbox{Pb--Pb}}
\newcommand{\AuAu}{Au--Au}
\newcommand{\sqrsNNnoEq}{$\sqrt{s_{\rm NN}}$}
\newcommand{\sqrsNN}{$\sqrt{s_{\rm NN}}=$}
\newcommand{\sqrs}{$\sqrt{s}=$}

\newcommand{\vtwo}{$v_2$}

\newcommand{\Npart}{$N_{\rm part}$}
\newcommand{\Ncoll}{$N_{\rm coll}$}

\newcommand{\W} {\ensuremath{W}}
\newcommand{\Z} {\ensuremath{Z}}
\newcommand{\D} {\ensuremath{D}}

\newcommand{\dndeta}       {\ensuremath{\mathrm{d}N_\mathrm{ch}/\mathrm{d}\eta}}

\newcommand{\ppbar}        {\mbox{$\mathrm {p\overline{p}}$}}

\newcommand{\beq}{\begin{equation}}
\newcommand{\eeq}{\end{equation}}
\newcommand{\beqn}{\begin{eqnarray}}
\newcommand{\eeqn}{\end{eqnarray}}

\usepackage{lineno}
\begin{document}
\begin{frontmatter}

\title{An overview of experimental results from ultra-relativistic heavy-ion collisions at the CERN LHC: bulk properties and dynamical evolution}

\author[YF]{Panagiota Foka}
\ead{yiota.foka@cern.ch}
\author[MAJ]{Ma{\l}gorzata Anna Janik\corref{CorrespondingAuthor}}
\ead{majanik@if.pw.edu.pl}

\address[YF]{GSI Helmholtzzentrum f\"ur Schwerionenforschung GmbH, Planckstra\ss e 1, 64291 Darmstadt, Germany}
\address[MAJ]{Faculty of Physics, Warsaw University of Technology, Koszykowa 75, 00710 Warsaw, Poland}

\cortext[CorrespondingAuthor]{Corresponding author}

\begin{abstract}
The first collisions of lead nuclei, delivered by the CERN Large Hadron Collider (LHC) at the end of 2010, at a centre-of-mass
energy per nucleon pair \sqrsNN\ 2.76 TeV, marked the beginning of a new era in ultra-relativistic heavy-ion physics.
Following the Run 1 period, LHC also successfully delivered \PbPb\ collisions at the collision energy \sqrsNN\ 5.02 TeV
at the end of  2015.
The study of the properties of the produced hot and dense strongly-interacting matter at these unprecedented energies is
experimentally pursued by all four big LHC experiments, ALICE, ATLAS, CMS, and LHCb.
This review presents selected experimental results from  heavy-ion collisions delivered during the first three years of the LHC operation focusing on the bulk matter properties and the dynamical evolution of the created system.
It also presents the first results from Run 2 heavy-ion data at the highest energy,
as well as from the studies of the reference pp and \pPb\ systems, which are an integral part of the heavy-ion programme.
\end{abstract}


\date{\today}

\end{frontmatter}

\section{Introduction}
\label{sec:Introduction}
The Standard Model, extensively tested by the high-energy physics community, 
and recently validated once more with the Higgs discovery,
predicts a series of phase transitions.
Within the Standard Model,
quantum chromodynamics (QCD), the theory of strong interactions, predicts
(and numerical calculations on the lattice confirm) 
that nuclear matter undergoes a phase transition (crossover) to a state of deconfined quarks and gluons,
at a critical temperature\footnote{In fact, it is a pseudo-critical temperature as 'lattice QCD' calculations indicate a crossover rather than a well defined phase transition \cite{Aoki:2006we,PseudoCritical}.} of about \mbox{160 MeV \cite{Borsanyi:2010cj}}, 
associated to an energy density of about 0.7 GeV/fm$^3$~\cite{Karsch:2000kv}. 
In addition, at about the same temperature,
chiral symmetry is expected to be (approximately) restored and quark masses are reduced from their large effective values in hadronic matter to their small bare ones.
While the Higgs mechanism is put forward to explain the mass generation of elementary particles,
the mass of composite particles, 
which is larger than the mass of their constituents due to their interactions, 
is to be understood by studying the predicted QCD phase transitions and the high-temperature phase of QCD.
Indeed, it was suggested that studying matter in such a phase, 
where quarks and gluons are no longer confined into hadrons, 
called the Quark-Gluon Plasma (QGP) \cite{Shuryak:1978ij},
one can gain insight into the basic features of QCD matter in its normal state, 
namely, confinement and chiral symmetry breaking \cite{Lee:1974ma}.
According to the Big Bang cosmology such a deconfined state of matter existed in the early Universe during the first few microseconds after its
creation  \cite{Boyanovsky:2006bf}, and it may also exist in the cores of neutron stars \cite{Alford:2013pma}.

\begin{figure}[hbt]
\centering
	\includegraphics[width=.39\textwidth,height=5.8cm]{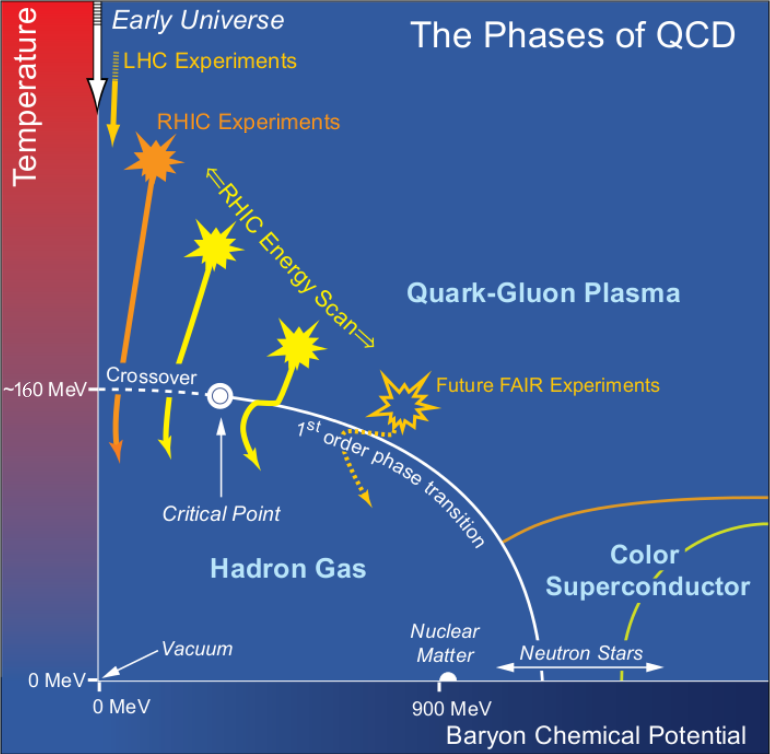}
	\includegraphics[width=.6\textwidth,height=5.8cm]{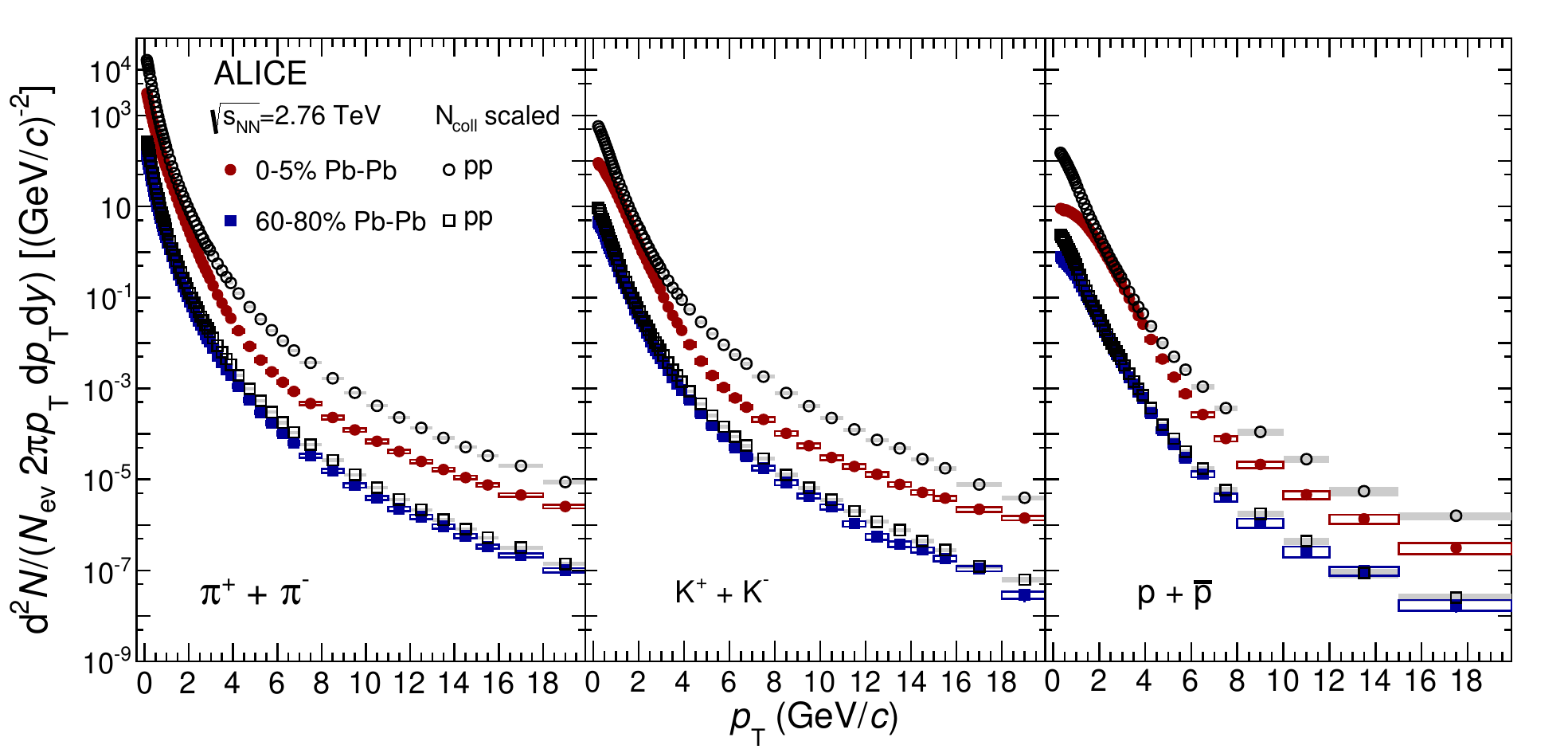}
	\caption{
	(Left)  The phase diagram of strongly interacting matter \cite{ALongRangePlan2008}. 
	(Right) Transverse momentum spectra of $\pi^{\pm}$, $\rm K^{\pm}$ and $\rm p$($\bar{\rm p}$) up to \mbox{\pt\ = 20 GeV/$c$} at mid-rapidity in \pp\ at $\sqrt{s}=2.76$ TeV, peripheral (60--80\%) and central (0--5\%) \PbPb\ collisions at \sqrsNN\ 2.76 TeV. Figure from \cite{Abelev:2014laa}.
   }			
	\label{fig:phaseDiagram} 
\end{figure}

Experimentally, collisions of heavy ions make it possible to create and study in the laboratory strongly-interacting matter under extreme conditions. 
Several facilities, worldwide, contribute to exploring the details of the QCD phase transitions, mapping out different domains of the QCD phase diagram, Fig.~\ref{fig:phaseDiagram}-left. Heavy-ion collisions at the top RHIC and LHC collider energies create deconfined matter characterized by small, almost vanishing, net-baryon densities and high temperatures compatible with lattice QCD calculations. Future facilities at FAIR and NICA are being constructed to study the other extreme of the phase diagram at high baryochemical potential and low temperature.

Historically, after pioneering experimental efforts in the 1970s at LBNL and JINR 
theoretical and experimental research advanced with increasing energy
at various laboratories from GSI SIS to BNL AGS and CERN SPS.
In 2000, an assessment of the SPS heavy-ion programme was presented \cite{Heinz:2000bk,CernPress} 
concluding that a new state of matter was produced in \PbPb\ central collisions at \sqrsNN\ 17.3 GeV, 
exhibiting some of the most important characteristics predicted for the QGP. 
Further progress of the heavy-ion programme at the BNL RHIC \cite{Arsene:2004fa, Adcox:2004mh} and at the CERN SPS \cite{Specht:2010xu} confirmed and refined the first SPS results.
The study of QCD at high temperatures 
entered a new era of precision measurements in 2010 with the startup of LHC 
delivering \PbPb\ collisions at the largest ever collision energies,
more than an order of magnitude larger than previously achieved. 

The experimental characterization of the QGP state should provide insight into the yet unexplained features of  QCD that are important to understand hadron and nuclear properties.
The study of heavy-ion collisions probes QCD in the non-perturbative domain
which calls for the development of effective theories. 
While numerical calculations on the lattice advanced considerably,
they have difficulties in addressing dynamical quantities, such as transport coefficients (for first attempts see \cite{Meyer:2007ic,Meyer:2007dy}).
Thus, in order to extract the properties of the produced matter,
phenomenology and Monte Carlo models have to be employed. 
Typically, at low and intermediate energies, hadronic transport models are used~\cite{Danielewicz:2002pu, Aichelin:1991xy, Petersen:2008kb, Buss:2011mx}, while at high energies hydrodynamics
is applicable \cite{Huovinen:2006jp, Gale:2013da}. 
In addition, approaches based on the combination of hydrodynamics and transport, are employed \cite{Petersen:2008dd}
to model the full evolution of the system.
 Lattice QCD methods are known to fail in several contexts, the most important of which are nonzero baryon density, the determination of inherently Minkowskian (real time) quantities, and systems out of thermal equilibrium. In these cases, remaining first principles methods include perturbation theory, relying on the weak coupling limit, as well as the gauge/gravity duality, which applies to a class of conformal field theories in their strongly coupled, large-$N_c$ limits. Both of these methods have been successfully applied to the determination of a large number of quantities relevant for heavy-ion physics. For a review, see e.g.~\cite{Brambilla:2014jmp}. 
Based on the comparison of models to the experimental data over a broad range of collision energies, it is possible to extract the following ranges for the macroscopic characteristics of the produced fireball 
(given in the system of units where $\hbar = c = 1$) 
\cite{Andronic:2014zha} (1) temperature: 100 -- 500 MeV, of the order of a million times the temperature of the
centre of the Sun (1 MeV$\approx 10^{10}$~K), (2) volume: (1--5) $ \cdot 10^3$ fm$^3$, (3) lifetime: (10 -- 20) fm/$c$ (about \mbox{(3 -- 6) $ \cdot$ $10^{-23}$ s}), (4)~pressure: 100 -- 300 MeV/fm$^3$ (1 MeV/fm$^3 \approx 10^{33}$ Pa), (5)~density: (1 -- 10) $ \cdot$ $\rho_{0}$ (normal nuclear matter density $\rho_{0} =0.17\ {\rm fm}^{-3} = 2.7$ $\cdot$ $10^{14}\ {\rm g/cm}^3)$.

To extract the properties of the produced matter different experimental observables are being optimized to probe the dynamical evolution of the system and characterize the different stages of the collision. 
At ultra-relativistic energies ($E/m >> 1$), during the collision of the two nuclei, dense gluon fields create a  strongly-interacting medium, which then rapidly expands and very quickly thermalizes~\cite{Muller:2013dea}. 
As the thermalized QGP continues to expand it cools down
till its temperature decreases below the critical temperature of the QCD phase transitions $T_{\rm c} \sim 160$~MeV, 
at which point it hadronises and converts into a hadron-resonance gas. At this moment, the composition of the produced particles  is approximately fixed, at a temperature called the chemical freezeout temperature (which is presumably close to the critical temperature of the phase transitions). 
After the chemical freezeout hadrons continue to interact. However, only their momentum distributions are modified 
as their relative energy is below the  inelastic production threshold. 
At kinetic freezeout, characterized by the kinetic freezeout temperature, 
the medium is so dilute that the final state hadrons cease interacting and decouple.
At this moment their momenta are fixed.
According to common lore, the initial state of a heavy-ion collision can be described using the so-called Color Glass Condensate (CGC) framework \cite{Kovner:1995ja,Kovchegov:1997ke,Krasnitz:1998ns,Krasnitz:1999wc,Lappi:2003bi,Lappi:2006fp}, featuring an overoccupied ensemble of soft gluons up to the so-called saturation scale, above which the states are practically unoccupied. The evolution of the system from this state towards hydrodynamic behaviour -- and ultimately full thermalization -- can today be described both at very weak and strong couplings, using effective kinetic theory and numerical holography, respectively. Both pictures show fast apparent hydrodynamization at time scales of order of 1 fm/$c$ or smaller~\cite{Chesler:2010bi,Kurkela:2015qoa}.

It is mostly from the final state hadrons measured in the experimental apparatus 
that one tries to deduce information about the initial state and the collision history
as it is not possible to directly detect the QGP. The most direct signal of the QGP state 
are photons shining through it, unaffected by strong interactions. 
The investigation of the creation of the QGP and the study of its properties is thus relying on appropriate experimental observables (so-called ``signatures'') and their comparison to models.
The characterization of the created partonic matter in terms of its initial conditions (eccentricity, volume, temperature, lifetime), equation-of-state (relating pressure and energy) and of its transport properties (viscosity and diffusion coefficients) 
and ultimately delineating the QCD phase diagram \cite{BraunMunzinger:2008tz} is the goal of a major ongoing research effort \cite{BraunMunzinger:2007zz, Jacak:2012dx, Muller:2013dea, Satz:2013xja, Schukraft:2013wba}.
Such studies have shown that \AuAu\ collisions at the top RHIC energies, at \sqrsNN\ 200 GeV, produce a deconfined strongly interacting medium which behaves like an almost perfect liquid, with minimal shear viscocity to entropy ratio, very close to the minimum theoretical limit, and it is opaque to hard probes, quenching energetic particles propagating through it.

At the LHC, the huge increase of energy, up to \sqrsNN\ 5.02 TeV for \PbPb\ collisions, 
is expected to 
provide more favourable conditions of  energy density and temperature  
that should lead to the creation of a denser, hotter, longer-lived medium.
Hence, one of the goals for the heavy-ion programme at LHC was to measure with increased precision the parameters which characterize this new state of matter, also making use of the particular strength of the LHC; namely, a powerful new generation of large acceptance, state-of-the-art, experiments, ALICE, ATLAS, CMS, and LHCb.
The advantage at the LHC and the most important impact of the increase of the collision energy is the large increase of the production of hard probes, giving access to a new class of observables, which is the focus of an  accompanying article also published in this journal \cite{P2}.

Experience from RHIC has shown that the bulk, macroscopic, properties of the produced matter can be well
described by hydrodynamics which therefore could be used to extrapolate their values to LHC energies. 
However, due to the huge energy jump  from RHIC to LHC (by a factor 10), 
its validity for the collisions at LHC
could by no
means be taken for granted. 
Therefore, the first task at LHC was to study typical observables that test the hydrodynamic description of the created system 
and probe its global properties, as discussed in this article.

The interpretation of the heavy-ion experimental data relies considerably on a systematic comparison with the same observables measured in proton--proton and proton--nucleus collisions as well as in collisions of lighter ions to study the system-size and energy-density evolution of the different experimental observables. In this way, the phenomena truly indicative of final-state effects due to the produced  deconfined medium can be distinguished from other contributions and initial-state effects characteristic of the initial nuclei. 
Therefore, the study of \pp\ and \pPb\ reference data at the same energy is integral part of the heavy-ion physics programme at LHC.
In particular, systematic comparisons and validation of the pp results with perturbative-QCD calculations is the first necessary step of such a programme.
In addition, comparisons of \pPb\ results to appropriate models are expected to disentangle
effects characteristic of cold nuclear matter (CNM) present at the initial state nuclei such as changes in the nuclear
parton distribution functions or hadronic final state interactions \cite{Andronic:2015wma,Kluberg:2009wc,Brambilla:2010cs}.
For such collisions, at high energies, and especially at the extreme LHC energies, it is important to determine the achieved conditions and, in particular, the energy density.
Thus, systematic studies of the small reference systems, pp and \pPb , offer, in addition, the possibility to investigate some basic physics questions like
the onset of collectivity and in more detail,
the number of interactions required between particles to treat a collection of them as a true medium as well as the conditions for the related transition from a microscopic to a macroscopic description in a relativistic quantum system.

To  systematically study heavy-ion collisions  certain experimental control parameters have been established and used routinely, for details see \cite{CasalderreySolana:2011us}. Examples of such parameters are the collision energy per nucleon pair, \sqrsNNnoEq\ (which can be varied by changing the beam energy), as well as  the overlap area of the colliding nuclei (which depends on the collision centrality or, alternatively, the size of the colliding nuclei).
Experimentally, 
specific measurements \cite{Abelev:2013qoq} are used to express collision centrality,
usually given as a percentage of the geometrical cross section.
An alternative way, commonly used, is via the number of participating nucleons, \Npart, which are the
nucleons involved in the formation of the fireball in the overlap area of the colliding nuclei \cite{Gosset:1976cy}.
\Npart\ is estimated as an average over a given centrality range using a model describing the geometry of the collisions, so-called Glauber model \cite{Miller:2007ri}.  
From the same model  \Ncoll\ can be estimated,
which is the number of single nucleon-nucleon collisions.
Central collisions of large, heavy ions are expected to provide ideal conditions of high-energy density and temperature to create a macroscopic system, the predicted hot and dense deconfined QGP medium~\cite{Lee:1974ma}.

The different  underlying physics processes are reflected in different ranges of the transverse momentum  (\Pt ) spectra (Fig.~\ref{fig:phaseDiagram}-right).
In a rough classification, three domains can be identified: low, intermediate and high \pt . 
At low \mbox{\pt\ $< 2$~GeV/$c$}, the bulk-matter dynamics is well
described by relativistic hydrodynamic models. Even at the high LHC energies, 
a large fraction of all particles (more than 95$\%$) are produced in this \Pt\
regime. The shape of the spectra of produced particles reflects the conditions at kinetic freezeout and the integrated particle yields reflect the conditions at
chemical freezeout.
At high \Pt\ $ > 8$~GeV/$c$, the spectrum is dominated by partons from hard processes (i.e. parton scatterings with large momentum transfer) interacting with the medium. 
Hard, high-\Pt\ observables are found to scale with \Ncoll , while soft observables at low \Pt\ scale with \Npart .  
Understanding the interplay of soft and hard processes, 
particularly important at intermediate \Pt\ range,
as well as the onset of hard processes, 
is relevant for understanding fundamental properties of QCD;
it remains, however, a theoretical challenge 
which will benefit from experimental input and is therefore a major task at LHC.

A further advantage at LHC is the very broad phase-space coverage achieved by the detector systems also in pseudorapidity $\eta$ ($\eta$ is associated to the polar angle of the momentum of the particle relative to the beam axis\footnote{$\eta=- {\rm ln} \left [ {\rm tan} \frac{\theta}{2}\right ]$, where $\theta$ -- polar angle; when the mass of the particles is negligible, $\eta$ is a good approximation of the rapidity  $y = \frac{1}{2}{\rm ln} \left ( \frac{E+p_L}{E-p_L} \right )$.}, with the remnants of the colliding nuclei flying at forward (backward) rapidities along the beam direction).
Furthermore, the LHC experiments use almost all known particle identification techniques and the possibility to discriminate different particle species can be used as an additional powerful parameter.
Measurements include a large variety of particles, light and strange hadrons, isolated photons, \Z, \W, \D, \jpsi\ (prompt and from $B$ decays), $\Upsilon$ as well as electrons and muons, with electro-weak bosons being identified for first time in heavy-ion collisions.

Overall, the high 
energies at LHC together with optimized experiments
open up the possibility for precision, multi-differential measurements over an extended phase space in \Pt\ and rapidity,  
also exploiting particle identification; in particular, the \Pt\ coverage, depending on the observables,
extends from almost zero (Fig.~\ref{fig:phaseDiagram}-right), up to, in some cases, hundreds of GeV$/c$.  
The simultaneous measurements of both soft and hard probes is mandatory to fully characterize the created system; however, due to space limitations we discuss in this paper a selection of low and intermediate \pt\ results and of hard processes in the accompanying article  \cite{P2}.

The first run of LHC, from 2009 to 2013, provided a wealth of measurements in pp collisions at $\sqrt{s}$  from 0.9 to 8 TeV,  \pPb\ collisions at \sqrsNN\ 5.02 TeV, and \PbPb\ collisions at \sqrsNN\ 2.76 TeV, also including ultra-peripheral  (photon-induced) nucleus-nucleus collisions. The second run already provided first measurements of \pp\ collisions at $\sqrt{s}$ = 13 TeV and \PbPb\ collisions at \sqrsNN\ 5.02 TeV with more to follow.

This article presents an overview of the current status of heavy-ion research at LHC focusing on the global bulk matter properties and dynamics of the created system. 
The presented results are mostly from  \PbPb\ collisions at \sqrsNN\ 2.76 TeV collected during the three years of LHC Run 1; we also present first results of Run 2 from \PbPb\ collisions at  \sqrsNN\ 5.02 TeV published at the time of writing. 
In addition, we discuss results from pp and \pPb\ collisions relevant to the soft observables. 
We present a similar review of results on hard probes in the same journal \cite{P2}.
Other reviews of LHC results and further references to the literature can be found in \cite{Andronic:2014zha,Brambilla:2014jmp,Armesto:2015ioy,Roland:2014jsa,Loizides:2016tew,Schukraft:2013wba, Muller:2012zq,Norbeck:2014loa}.

\section{Global event properties} \label{sec:globalEventProperties}
\begin{figure}[ht]
	\begin{center}
		\includegraphics[width=0.47\textwidth,height=6.15cm]{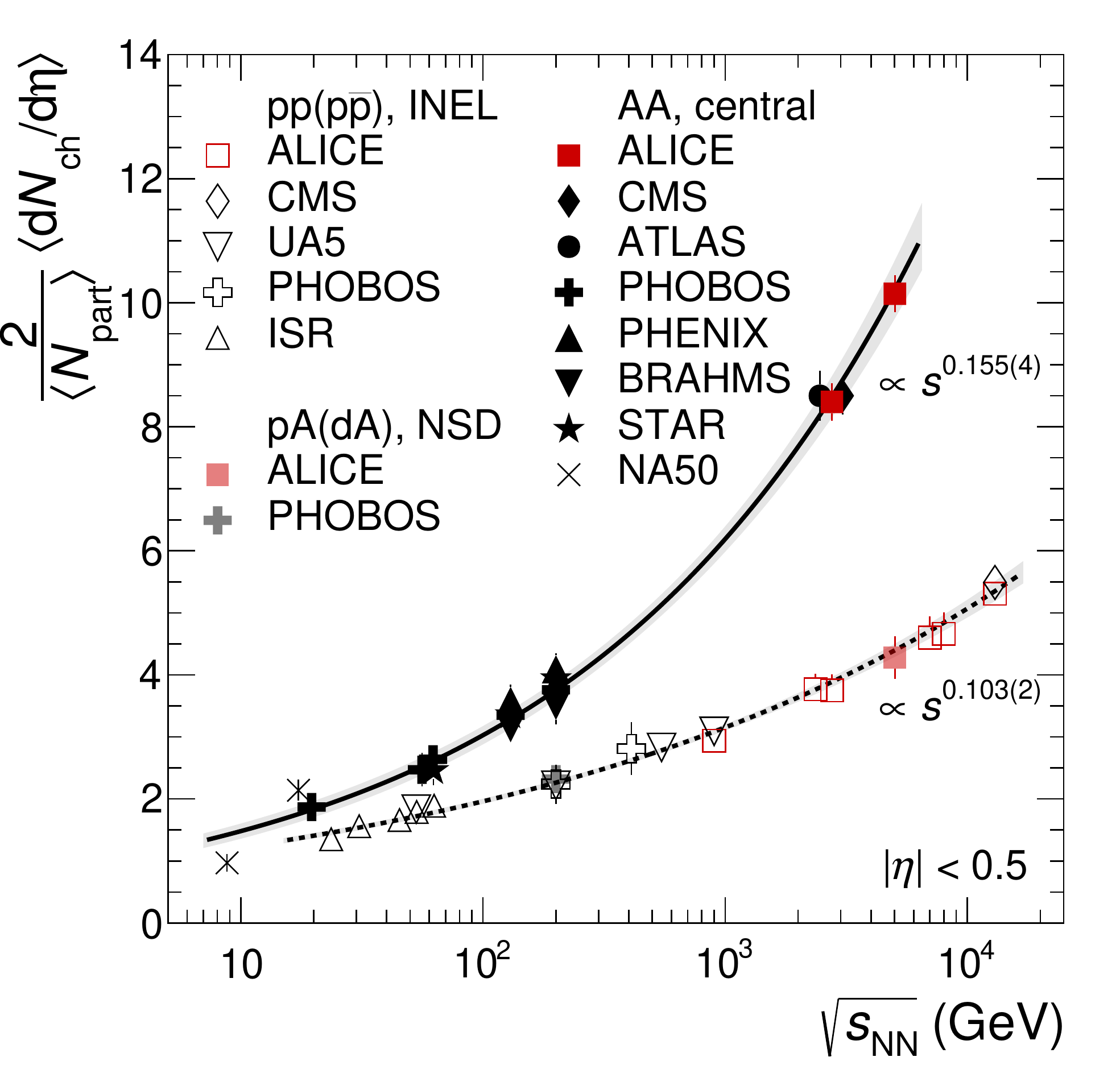}
		\includegraphics[width=0.51\textwidth,height=6.0cm]{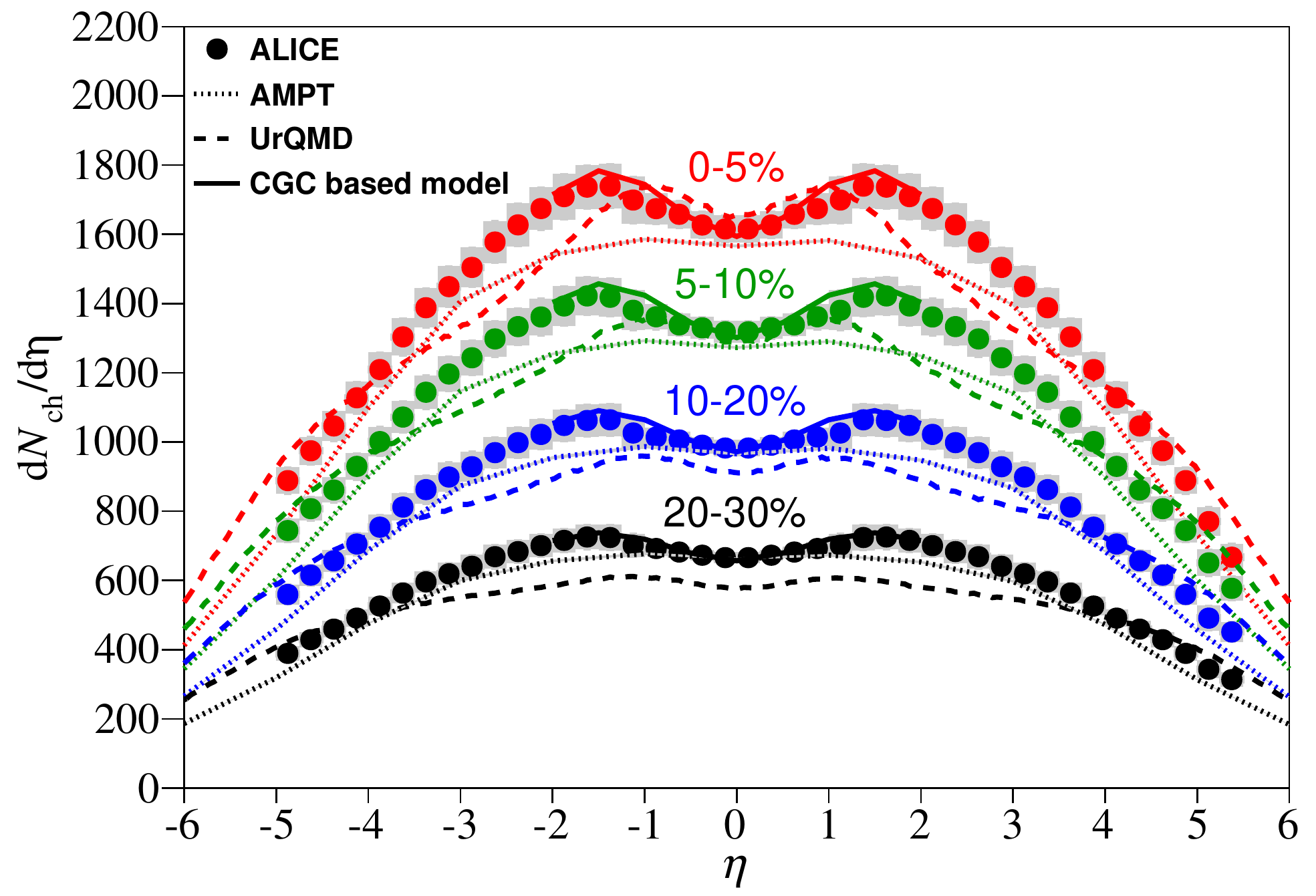}
		\put(-105,155){\PbPb \sqrsNN 2.76 TeV}
	\end{center}
	\caption{
		(Left) Charged-particle pseudorapidity density per participant pair as a function of collision energy, 
		for central \PbPb\ collisions \cite{Aamodt:2010cz,ATLAS:2011ag,Chatrchyan:2011pb,Abreu:2002fw,Adam:2015ptt}, and \AuAu\ collisions \cite{Bearden:2001xw,Bearden:2001qq,Adcox:2000sp,Alver:2010ck,Abelev:2008ab}, as well as for inelastic \pp\ and \ppbar\ events \cite{Adam:2015gka,Khachatryan:2015jna,Adam:2015pza}, and pA and d--A non-single diffractive collisions~\cite{Aamodt:2010cz,Back:2003hx}. The power-law dependence for AA and pp collisions is shown by solid and dashed lines, respectively. The \PbPb\ measurements from CMS and ATLAS at \sqrsNN\ 2.76 TeV are slightly shifted in \sqrsNNnoEq\ for visibility. Figure from~\cite{Adam:2015ptt}.
		(Right) Charged-particle pseudorapidity density distribution, \dndeta, for \PbPb\ collisions at \sqrsNN\ 2.76~TeV for four centrality classes, compared to model predictions~\cite{ALbacete:2010ad,Mitrovski:2008hb,Lin:2004en,Xu:2011fi}. Figure from~\cite{Abbas:2013bpa}.
	}
	\label{fig:Global1}
\end{figure}

\paragraph{Charged hadron multiplicity}
The number of produced particles (multiplicity) is an important property of the collisions related to the initial energy density and collision geometry and it is sensitive to the interplay between particle production from hard and soft processes as well as to coherence effects between individual nucleon-nucleon scatterings. With an increase of collision energy it is expected that the role of hard processes increases.
Because particle multiplicity cannot be currently calculated from first principles, in particular at low \pt, 
experimental measurements provide crucial input to models describing particle-production mechanisms.
A summary of  charged-particle pseudorapidity density (charged particle multiplicity per rapidity unit, d$N_{\rm ch}/{\rm d}\eta$) per participant pair from pp, pA, and AA collisions, measured as function of the collision energy, is shown in Fig.~\ref{fig:Global1}-left \cite{Adam:2015ptt}, including data at the highest available energies from \pp\ at $\sqrt{s}=13$~TeV and \PbPb\ at \sqrsNN\ 5.02~TeV. The dependence of this quantity on the collision energy follows a power law behaviour, $s_{NN}^\alpha$, with $\alpha = 0.1$ for the elementary systems pp and pA, while the AA data show a much steeper dependence, best described with $\alpha = 0.16$. 
The change from RHIC to LHC is related to a possible enhanced contribution of hard processes at the higher energies. The 20\% relative increase observed between the \PbPb\ measurements at \sqrsNN\ 2.76 TeV and 5.02 TeV is in agreement with established trends.  The steeper rise of particle production in AA with respect to elementary pp and pA shows that heavy-ion collisions cannot be  described as an independent superposition of single nucleon-nucleon interactions. Such measurements are employed to constrain theoretical models and help disentangle different production processes \cite{ATLAS:2011ag,Muller:2012zq,Adam:2015ptt}. 

Figure~\ref{fig:Global1}-right shows the measurement of the charged-particle pseudorapidity distribution~\cite{Abbas:2013bpa,Aamodt:2010cz,ATLAS:2011ag,Chatrchyan:2011pb,Adam:2015ptt} over a broad $\eta$ range ($-5<\eta<5$) from \PbPb\ collisions at \sqrsNN\  2.76~TeV, compared to theoretical calculations~\cite{ALbacete:2010ad,Mitrovski:2008hb,Lin:2004en,Xu:2011fi}. 

It is a challenge for the models to describe the shape and values of the distribution in the whole $\eta$ range. However, the Color Glass Condensate  model \cite{Iancu:2003xm}, based on gluon saturation at the initial state, can describe the data in the region where its application is valid. This measurement also allows us to extract the number of charged particles produced in a collision, which for the 0--5\% central \PbPb\ events at \sqrsNN\ 2.76~TeV obtained from the integration of the distribution over the whole $\eta$ range reaches $N_{\rm ch}=17165\pm 772$~\cite{Abbas:2013bpa},  and measured in $|\eta|<0.5$ reaches $1601\pm60$ \cite{Aamodt:2010cz}.

\begin{figure}[ht]
\begin{minipage}{0.49\textwidth}
	\hspace{0.4cm}
	\includegraphics[height=5cm,width=7.5cm]{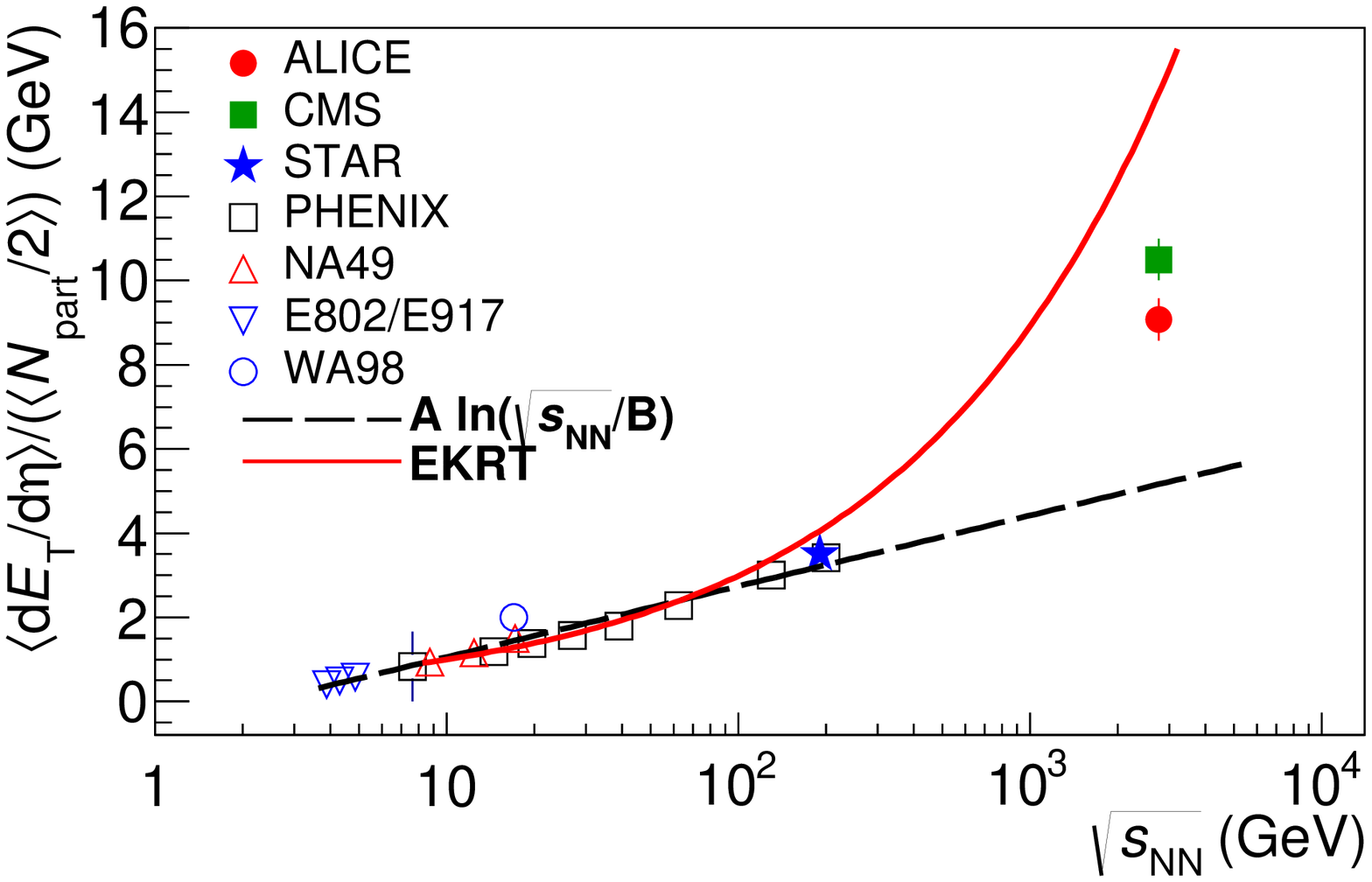}
	\put(-230,130){(a)}
	\hspace{0.4cm}
	\includegraphics[height=5cm,width=7.5cm]{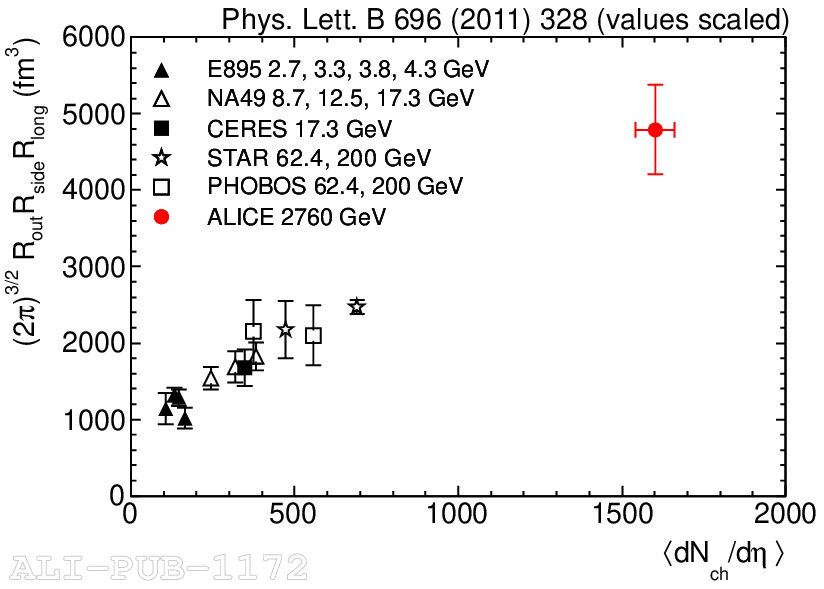}
	\put(-230,127){(c)}
\end{minipage}
\begin{minipage}{0.49\textwidth}
	\includegraphics[height=9.7cm]{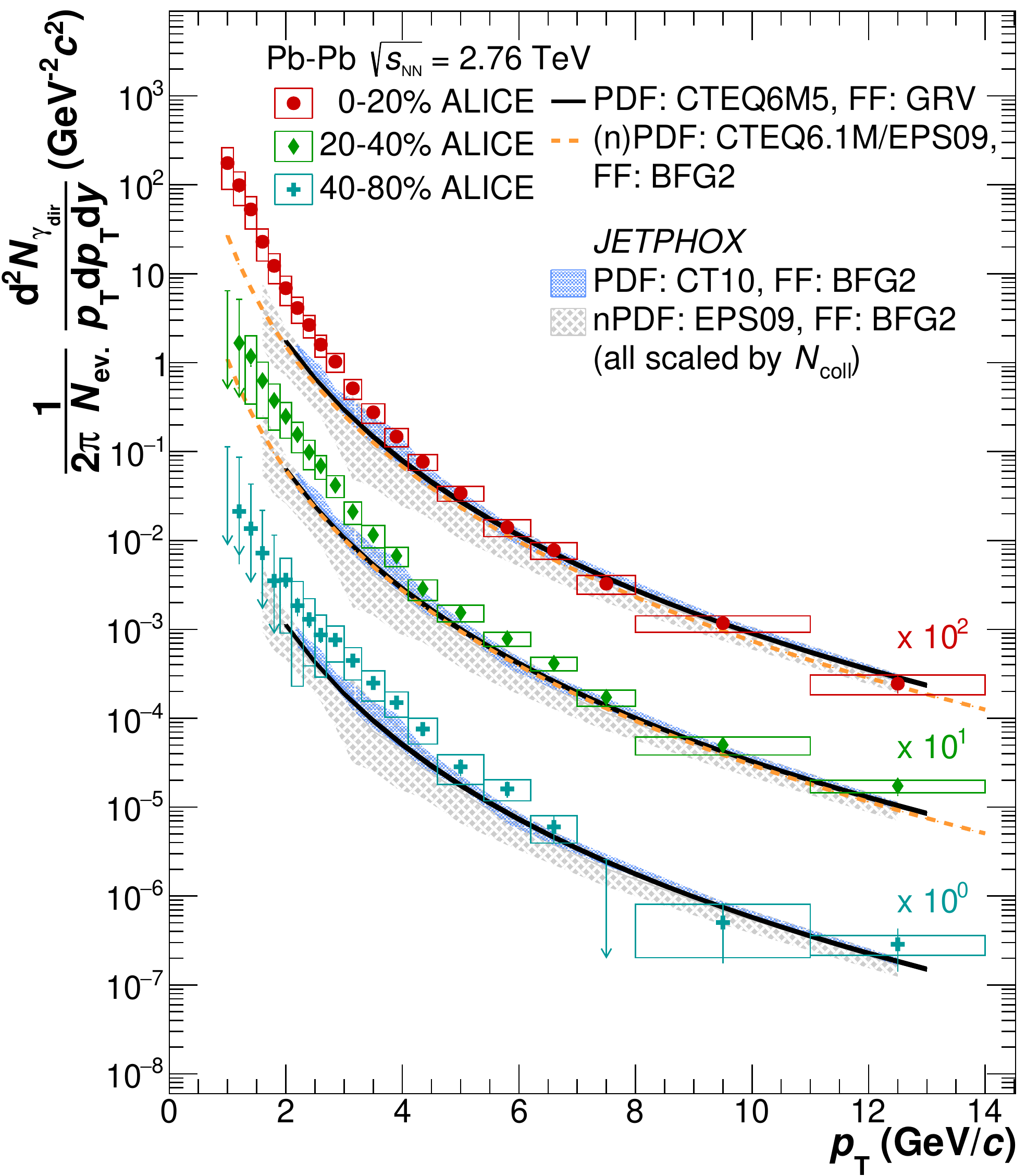}
	\put(-248,265){(b)}
	\put(-188,35){photons}
\end{minipage}
	\caption{(a)  Transverse energy pseudorapidity density 
	per participant pair
	for central  AA collisions (0--7\% centrality for NA49 and 0--5\% for all other experiments) at midrapidity as a function of collision energy. Figure from \cite{Adam:2016thv}.
	(b) Direct photon  $p_{\rm T}$ spectrum for \PbPb\ collisions at $\sqrt{s_{\rm NN}}=2.76$ TeV~\cite{Adam:2015lda}, compared to NLO pQCD calculations. Figure from~\cite{Adam:2015lda}.
	(c) Effective ``volume" of the particle emitting source at kinetic freezeout calculated as a product of the pion HBT radii for heavy-ion collisions at the LHC (centrality 0--5\%) and lower energies \cite{Aamodt:2011mr}, shown  
	as a function of charged-particle pseudorapidity density.
	Figure from~\cite{Aamodt:2011mr}.
	}
	\label{fig:photons}
\end{figure}

\vspace{-0.2cm}
\paragraph{Transverse energy} The transverse energy pseudorapidity density was measured for $\sqrt{s_{\rm NN}}=2.76$~TeV \PbPb\ data first by CMS~\cite{Chatrchyan:2012mb}, and recently by ALICE \cite{Adam:2016thv}, Fig.~\ref{fig:photons} (a).
This quantity is used to estimate the energy density within the Bjorken hydrodynamic model \cite{Bjorken:1982qr}, which is based on geometrical considerations (longitudinal, transversely isotropic expansion of the created medium). In that case the energy density is $\epsilon=\left ( \mathrm{d}E_{\rm T}/\mathrm{d}\eta \right )|_{\eta=0}/\left ( A \times \tau_0 \right )$, where $A$ is the overlap area of the colliding nuclei and $\tau_0$ is the parton-formation time~\cite{Adler:2004zn}. The ALICE results are consistent with the ones of CMS for the 10--80\% centrality range; however,  the ALICE  measurement yields a lower transverse energy density for the 0--5\% central collisions,  of  $\epsilon$ = 12.3$\pm$1 GeV/fm$^3$, compared to the CMS measurement  $\epsilon$ = 14~GeV/fm$^3$. Therefore, at the LHC, 
assuming $A=\pi\times(7\ \rm{fm})^2$ and $\tau_{0}=1$ fm/$c$,  
the energy density is in the range $\epsilon$ = 12--14~GeV/fm$^3$  for the initial stage of central \PbPb\ collisions at $\sqrt{s_{\rm NN}} = 2.76$~TeV, roughly 2--3 times higher than that
reported at RHIC \cite{Arsene:2004fa,Adcox:2004mh,Back:2004je,Adams:2005dq,Adler:2004zn} and 
well above the critical energy density predicted for the phase transition, which is about 0.7 GeV/fm$^3$~\cite{Karsch:2000kv}.

\vspace{-0.3cm}
\paragraph{Initial temperature}
The initial temperature of an equilibrated QGP state can be estimated by studying electromagnetic probes, 
 which do not interact strongly with the medium and therefore carry information from the early stages of the collision \cite{Carminati:2004fp}. A recent discussion of theoretical aspects of photon production in high-energy nuclear collisions can be found in \cite{Ghiglieri:2013gia}.
The measurement of direct photons\footnote{Direct photons are photons not originating from hadron decays. They may originate from different stages of the collision, i.e. direct prompt photons coming from the initial hard parton scatterings, direct thermal photons originating from the QGP state.} at $\sqrt{s_{\rm NN}}=2.76$~TeV Pb--Pb collisions is presented in Fig.~\ref{fig:photons} (b). The direct-photon spectrum was extracted by subtracting the contribution from particle decays to photons~\cite{Adam:2015lda}. The spectrum is compared to 
NLO pQCD calculations for pp collisions scaled by the number of binary collisions~\cite{Klasen:2013mga,Gordon:1993qc,Vogelsang:1997cq,Paquet:2015lta,Ghiglieri:2013gia},
which 
describe well the photon spectrum for $p_{\rm T}>5$ GeV/$c$. 
However, the excess below 2~GeV/$c$, not described by these calculations, can be attributed to thermal photons. 
At the LHC, the inverse slope parameter extracted  from an exponential fit at the low-\pt\  range $0.8<$\pT\ $<2.2$~GeV/$c$, an ``effective temperature," is $T_{\rm eff}=297 \pm 12({\rm stat})\pm 41({\rm syst})$~MeV.
This value is well above the temperature expected  for the phase transition  (about 160 MeV) and about 40$\%$ higher than the one reported at RHIC \cite{Adare:2008ab}. 
However, measurements of thermal photons are notoriously demanding \cite{Wilde:2012wc} and  important considerations have to be taken into account for the interpretation of these measurements~\cite{Abelev:2012cn}, i.e. advanced theoretical analyses of 
the data \cite{Klasen:2013mga} consider blue-shift Doppler corrections  taking into account the radially expanding medium \cite{Shen:2013vja}. Such corrections have been accounted for in recent model calculations involving hydrodynamical evolution of the QGP~\cite{vanHees:2014ida,Chatterjee:2012dn,Paquet:2015lta,Linnyk:2015tha}. The initial temperatures $T_{\rm init}$ assumed in these models vary between 350 and 700~MeV.

\vspace{-0.2cm}
\paragraph{Chemical freezeout}

Assuming the simple statistical nature of the hadronization process and that the created medium (with partonic degrees of freedom) is in thermodynamic equilibrium, thermal (statistical) models based on the grand canonical ensemble can be successfully employed to describe the particle abundances~\cite{BraunMunzinger:2003zd,BraunMunzinger:2003zz}. 
Within the framework of such models, 
the conditions at chemical freezeout (where particle composition is fixed) 
can be determined from the measured particle yields,
and are described by the fit parameters corresponding to the chemical freezeout temperature
($T_{\rm ch}$) and baryochemical potential ($\mu_{\rm B}$).
Comparisons of the measurements from \PbPb\ collisions at \sqrsNN\ 2.76 TeV~\cite{Abelev:2013vea,Abelev:2013xaa,ABELEV:2013zaa,Knospe:2013tda,Sharma:2016vpz,Guerzoni:2016xdw} with
statistical models~\cite{Torrieri:2004zz,Wheaton:2004qb,Kisiel:2005hn,Andronic:2008gu,Petran:2013dva,Cleymans:2006xj}
show that the best agreement
is reached at vanishing baryochemical potential
$\mu_{\rm B}\approx 1$~MeV, and at a chemical freezeout temperature of 
$T_{\rm ch} \approx 156$~MeV.
This is lower than the value $T_{\rm ch} \approx 164$~MeV predicted based on extrapolations from RHIC energies before the LHC start-up~\cite{Andronic:2007vh}. 
This deviation is due to an overestimate of the proton yield and several
possible explanations have been suggested in~\cite{Andronic:2008gu,Steinheimer:2012rd,Adare:2013oaa,Szymanski:2012qu,Becattini:2012xb,Endrodi:2011gv,Kaczmarek:2011zz,Cai:2013pda,Adam:2015vja}
including the effects of possible large baryon-antibaryon annihilation in the late hadronic phase.

\vspace{-0.2cm}
\paragraph{Kinetic freezeout}
The kinetic freezeout can be described by hydrodynamics-inspired blast-wave models~\cite{Back:2004je,Schnedermann:1993ws,Abelev:2008ab}.  Such models allow us to experimentally extract the associated temperature $T_{\rm kin}$ and the average transverse flow velocity $\langle \beta_{T} \rangle$ (reflecting the transverse expansion of the medium) from the low-\pt\  spectra of identified charged hadrons.  The spectra of $\pi$, $K$, and $p$ were measured at the LHC~\cite{Abelev:2012wca,Abelev:2014laa} from down to almost zero \pt\ and up to \mbox{\pt\ = 20 GeV/$c$} at midrapidity, as shown in Fig.~\ref{fig:phaseDiagram}-right. Because of the common radial expansion velocity, the \Pt\ spectra of different particle species reflect the differences of their masses which results in a characteristic mass ordering.  
The mean transverse momentum of the spectra is higher than in central \AuAu\ collisions at $\sqrt{s_{\rm NN}} = 200$~GeV at RHIC~\cite{Abelev:2008ab,Adler:2003cb}, which suggests stronger collective radial flow. 
A blast-wave fit 
yields a kinetic freezeout temperature $T_{\rm kin} = 96 \pm 10$~MeV, similar to the one at RHIC, and a radial flow velocity $\langle \beta_{\rm T} \rangle = 0.65 \pm 0.02$, about 10$\%$ higher than that at RHIC. The data are in good agreement with hydrodynamic model calculations~\cite{Shen:2011eg,Karpenko:2011qn,Karpenko:2012yf,Bozek:2011ua,Werner:2012xh} that include viscuous corrections and rescatterings at the hadronisation phase.

\vspace{-0.2cm}
\paragraph{Size and lifetime of the QGP medium}
The space-time evolution of the expanding fireball can be inferred
using identical pion interferometry techniques, also known as Hanbury-Brown Twiss (HBT) correlations
\cite{Lisa:2005dd},
and is found to be well described by calculations of hydrodynamical models  \cite{Adam:2015vna}.
At LHC, the system created in the 5$\%$ most 
central \PbPb\ collisions at \sqrsNN\ 2.76 TeV
has a ``homogeneity volume" at kinetic freezeout (when strong
interactions cease) of $\sim$5000~fm$^3$, about twice as large as the one
measured at RHIC, see Fig.~\ref{fig:photons} (c). 
Combining the world measurements, a linear increase of the extracted volume as a function of charged-particle multiplicity is observed. 
The femtoscopic technique also allows us to extract the decoupling time, because  the size of the particle emitting region is inversely proportional to the gradient of the velocity of the expanding medium.  The decoupling time extracted by ALICE is of the order of 10 fm/$c$ \cite{Aamodt:2011mr}, 30\% larger than the one obtained at RHIC.

\section{Angular correlations and fluctuations} \label{sec:particlecorrelations}


The collective behaviour of the QGP medium, apparent in all measured soft observables probing its bulk properties and well described by hydrodynamics (see Sec.~\ref{sec:globalEventProperties}), yields particle correlations between final-state particles.
In addition, there are numerous other sources of correlations, e.g. jets, quantum statistics or Coulomb effects, conservation laws, decays of unstable particles. 
A powerful way to investigate them is the method of two-particle angular correlation in relative pseudorapidity ($\Delta\eta$) and azimuthal angle ($\Delta\varphi$) space. 
Each one of the listed correlation sources produce a different shape in the $\Delta\eta\Delta\varphi$ distribution and contributes to the global correlation landscape. 
For example, quantum statistics HBT effects can be seen at  $(\Delta\eta,\Delta\varphi)\sim(0,0)$ at low \pt , while momentum conservation appears as a $- {\rm cos}(\Delta\varphi)$ shape underlying the measured distribution.
The information about all those contributions allows us to study the intrinsic collective behaviour of the medium. Moreover, it is further exploited by ALICE using identified particles (pions, kaons, protons)   \cite{ABELEV:2013wsa,MJthesis} as well as by CMS using hadron-$K^0$ and hadron-$\Lambda$ correlations \cite{Khachatryan:2014jra}.
By studying the relative contributions of the different components in $\Delta\eta\Delta\varphi$ space 
one can gain insight in the interplay of the different correlation sources 
and hence on the details of the dynamic properties of the medium.

\begin{figure}[htbp]
	\begin{center}
		\includegraphics[width=0.25\textwidth]{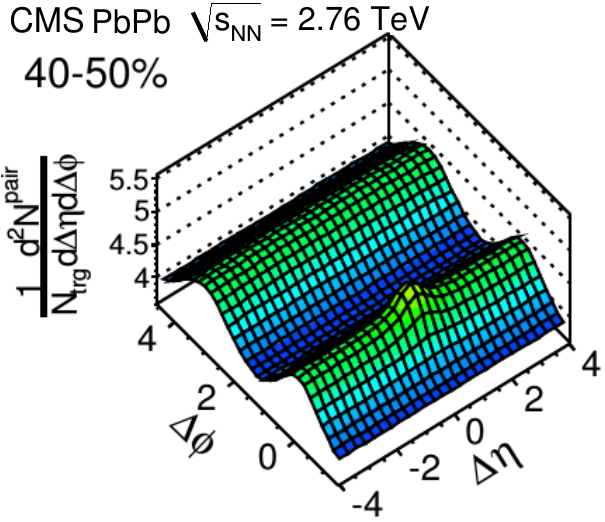}
		\put(-115,105){(a)}
		\includegraphics[width=0.49\textwidth]{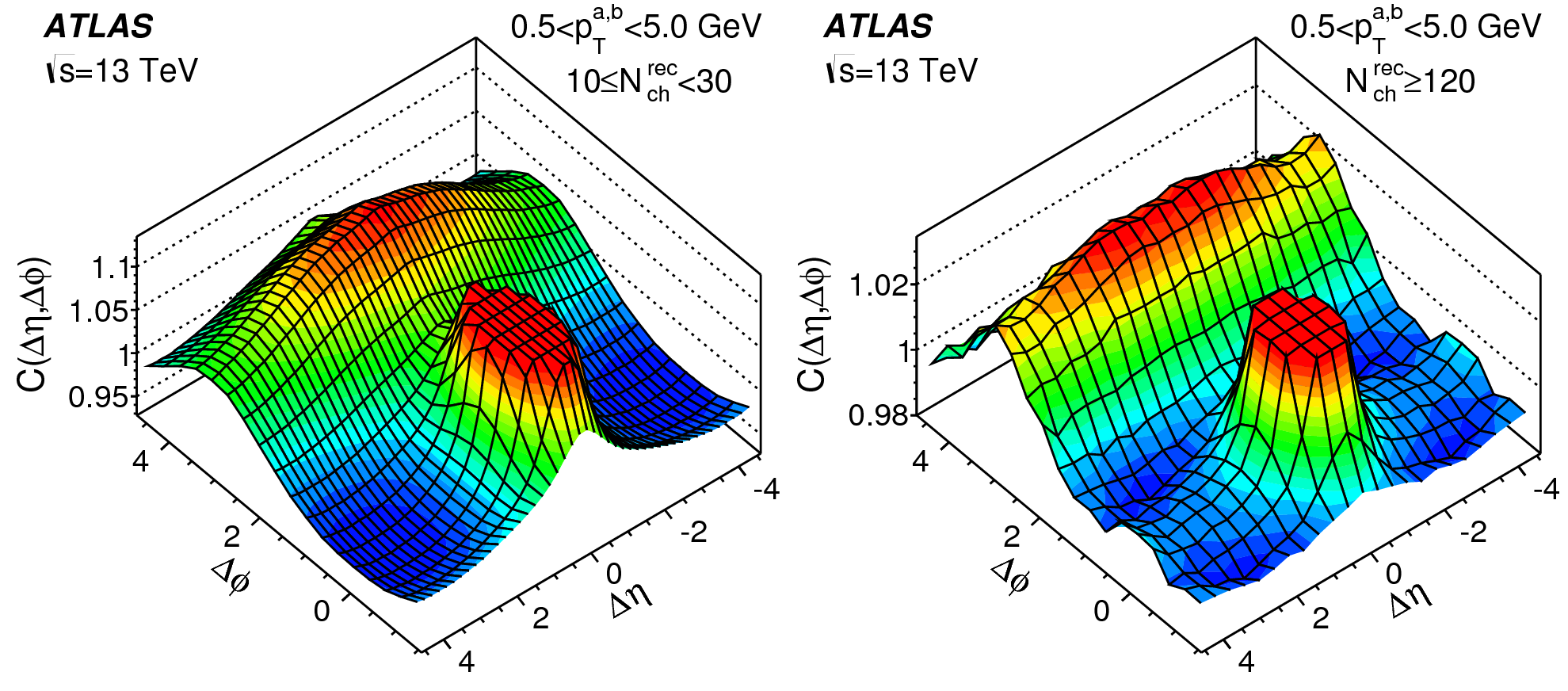}
		\put(-225,105){(b)}
		\put(-108,105){(c)}
		\includegraphics[width=0.25\textwidth]{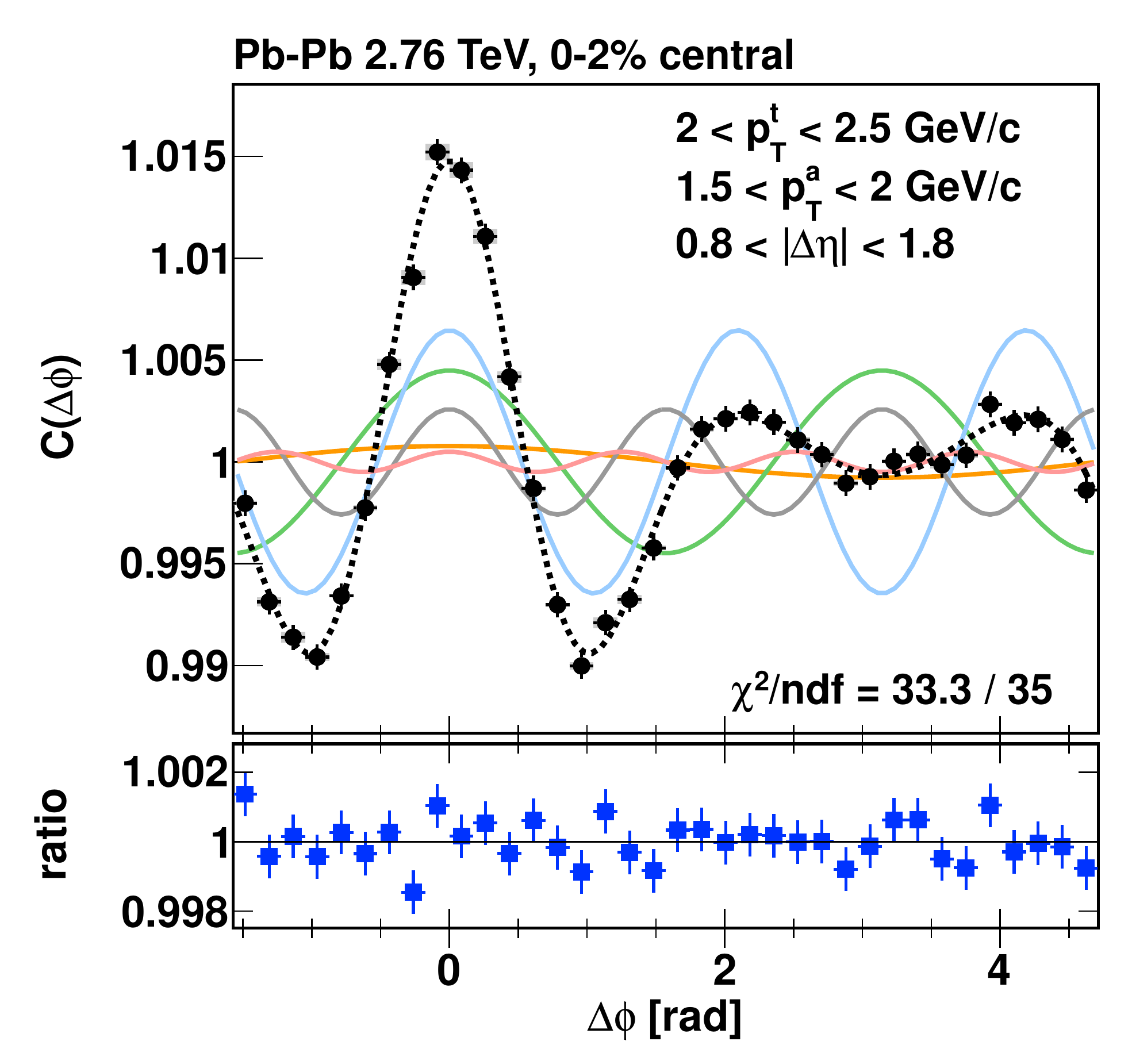}
		\put(-110,105){(d)}
	\end{center}
	\caption{(a) Two-dimensional \detadphi\ per-trigger-particle associated yield of charged hadrons 
		for \PbPb\ collisions at $\sqrt{s_{NN}}=$ 2.76 TeV.
		Figure from  \cite{Chatrchyan:2012wg}.
		(b--c) 
		Two-particle \detadphi\ correlation functions in $\sqrt{s}=$ 13 TeV pp collisions for charged particles in 0.5 $<$ \pT\ $<$ 5.0 GeV for two different multiplicity classes.
		The near-side peak is truncated  to better display the surrounding structure. 
		Figure from \cite{Aad:2015gqa}.
		(d) Two particle $\Delta\varphi$ correlation for particle pairs at $|\Delta \eta| > 0.8$, with the Fourier harmonics  $n=1$ to 5 superimposed in colour. Their sum is shown by the dashed curve. The ratio of data to the $n \le 5$ sum 
		is shown in the lower panel. 
		Figure from  \cite{Aamodt:2011by}. 
	}
	\label{fig:lowptcorr1}
\end{figure}

The \detadphi\
angular distributions, first measured in Au--Au collisions at RHIC~\cite{Alver:2009id,Abelev:2009af}, revealed two important features: a peak around $(\Delta\eta,\Delta\varphi)=(0,0)$, usually referred to as the ``near-side peak", originating primarily from jets, and prominent long-range correlations (referred to as the ``ridges"), extended over several units of rapidity and centered in azimuth around $\Delta\varphi=0$ (``near-side") and $\Delta\varphi=\pi$ (``away-side"). The near-side ridge becomes stronger with increasing centrality and has mostly been associated with collective phenomena. Analogous structures were observed in \PbPb\ collisions at the LHC  \cite{Aamodt:2011by,Chatrchyan:2011eka,Chatrchyan:2012wg}, as shown in Fig.~\ref{fig:lowptcorr1} (a).

Surprisingly, similar structures were observed in the ``small reference systems" at LHC. 
First,
a near-side ridge was observed in high-multiplicity pp collisions by CMS\footnote{Result sometimes referred to as ``the first LHC discovery" \cite{Schukraft:2013wba}.} and then by ATLAS \cite{Khachatryan:2010gv,Khachatryan:2015lva,Aad:2015gqa} at $\sqrt{s}=$~2.76 TeV, 7 TeV and 13 TeV pp data, 
followed up by similar results from  \pPb\ collisions at \sqrsNN\ 5.02  TeV~\cite{CMS:2012qk,Abelev:2012ola,Aad:2012gla,Aad:2014lta,Adam:2015bka}. 
In these systems a clear evolution from the absence of a near-side ridge correlation in low-multiplicity collisions to a prominent structure present at high multiplicities can be seen, see Fig.~\ref{fig:lowptcorr1} (b--c), while the jet-peak correlation remains roughly similar~\cite{Abelev:2014mva}.
ALICE results \cite{Abelev:2012ola} from \pPb\ collisions revealed that the near-side ridge is accompanied by an essentially identical away-side structure.
Recent results of ATLAS on \pp\ collisions  \cite{Aad:2015gqa}, implementing  
a sophisticated analysis method to study the multiplicity dependence of such effects, 
found the near-side ridges already to be present at very low-multiplicity classes; a smooth evolution of the magnitude of the effects is observed. 
There are two main possible explanations of this phenomenon under debate. The first one is based solely on hydrodynamics~\cite{Bozek:2012gr,Shuryak:2013ke,Bzdak:2013zma,Werner:2010ss,Gavin:2008ev}, while the second one originates from the gluon saturation picture of the initial state of the collision as implemented in CGC~\cite{Kovner:2010xk,Levin:2011fb,Dumitru:2010iy,Dusling:2012iga,Dusling:2012wy,Altinoluk:2015uaa}. 
For other possible explanations one can refer to~\cite{Hwa:2008um,Bjorken:2013boa,Shuryak:2013sra,Andres:2014bia}.

Assuming that the long-range correlations (for large $\Delta\eta$) 
are mostly due to collective phenomena, 
which induce correlations of practically all particles created in the collisions,
we can quantify them by extracting ``flow coefficients" (see below for detailed description) 
from the Fourier decomposition of the projections of the measured two-dimensional correlation 
into $\Delta\varphi$ at large $\Delta\eta$.  
Since non-flow effects usually induce only short-range correlations (at small $\Delta\eta$), 
they are expected to be strongly suppressed applying this procedure.
As an example, the projection of the two-particle correlation function $C(\Delta\varphi, |\Delta\eta| > 0.8)$ with the extracted Fourier harmonics is shown in Fig.~\ref{fig:lowptcorr1} (d).
At LHC, such coefficients have been extracted from the 
$\Delta\varphi$ distributions for all measured systems \PbPb , \pPb\, and pp; their interpretation and possible collective origin in the ``small systems" is the subject of intense systematic investigations.



In heavy-ion collisions collective phenomena have been associated 
with the presence of a macroscopic strongly-interacting medium.
The origin of hydrodynamic flow, in non-central heavy-ion collisions 
is the presence of anisotropic pressure gradients, 
developed in the overlap region of the two colliding nuclei 
(due to reinteractions among the produced medium constituents and/or produced final state particles), 
which then transform the initial spatial anisotropy into an observed momentum anisotropy, 
leading to an anisotropic particle distribution d$N$/d$\varphi$.
The azimuthal angle anisotropies can be quantified by applying 
a Fourier decomposition to the measured distribution~\cite{Voloshin:1994mz}.
Flow coefficients can be extracted from $v_{\rm n} = \langle \cos \big [n(\varphi - \Psi_{n}) \big] \rangle \,$,
where $\Psi_{\rm n}$ is the azimuthal angle of the symmetry plane of the overlap region, and $\varphi$ is the particle's azimuthal angle\footnote{This equation in valid for $n=1,2,...$. The zero order component $v_0$ is the average transverse velocity of the system’s collective radial expansion (averaged over all azimuthal angles).}.  The angle-averaged (isotropic) component $v_0$ is referred to as radial flow, the $v_{1}$ coefficient is known as directed flow, while
the second, $v_2$, Fourier coefficient, is the elliptic flow. 
Since the overlap region, in non-central heavy-ion collisions, has an approximately ellipsoidal shape, the dominant flow coefficient is $v_2$; however, other harmonics are also present.
The anisotropic flow 
is sensitive to the initial geometry of the overlap region as well as to the equation of state of the system and its transport properties. 

The presence of $v_2$ in AA collisions 
is one of the key observables indicating the creation of a deconfined medium that induces correlations
in contrast to a superposition of independent nucleon--nucleon collisions 
where no such azimuthal correlations are expected.

Detailed analysis of azimuthal correlations has shown that
the higher flow harmonics, initially assumed to be negligible due to symmetry reasons, are also present 
with measurements extending up to $v_6$~\cite{ATLAS:2012at,ALICE:2011ab,Chatrchyan:2013kba}.
They arise
due to the statistical nature of single nucleon-nucleon collisions which leads to irregular fluctuations of the initial energy-density profile of the colliding nucleons~\cite{Alver:2010gr, ALICE:2011ab, ATLAS:2012at, Chatrchyan:2013kba}.
These irregularities affect the initial pressure gradient and energy distribution and cause fluctuations of the direction and magnitude of the elliptic flow at an event-by-event level~\cite{Sorensen:2010zq,Alver:2010gr}.
Their measurement is particularly important to access and constrain the initial conditions of the collision, 
that play a key role in the determination of the $\eta/s$.

\begin{figure}[t!]
	\begin{center}
\begin{minipage}{0.49\textwidth}
\hspace{0.3cm}	
	\includegraphics[width=\textwidth]{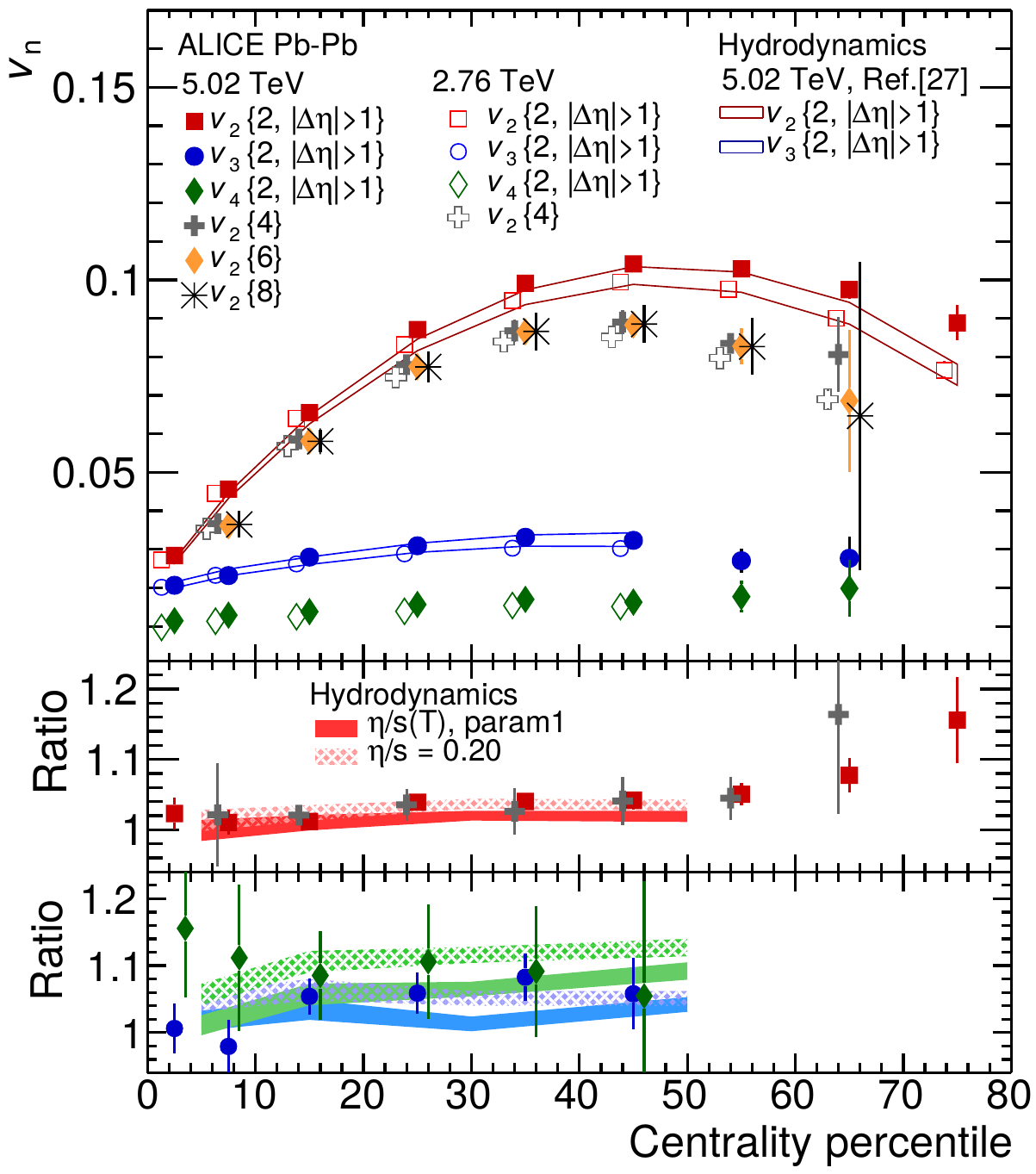}
	\put(-242,250){(a)}
\end{minipage}	
\begin{minipage}{0.49\textwidth}
\center
	\includegraphics[width=0.8\textwidth]{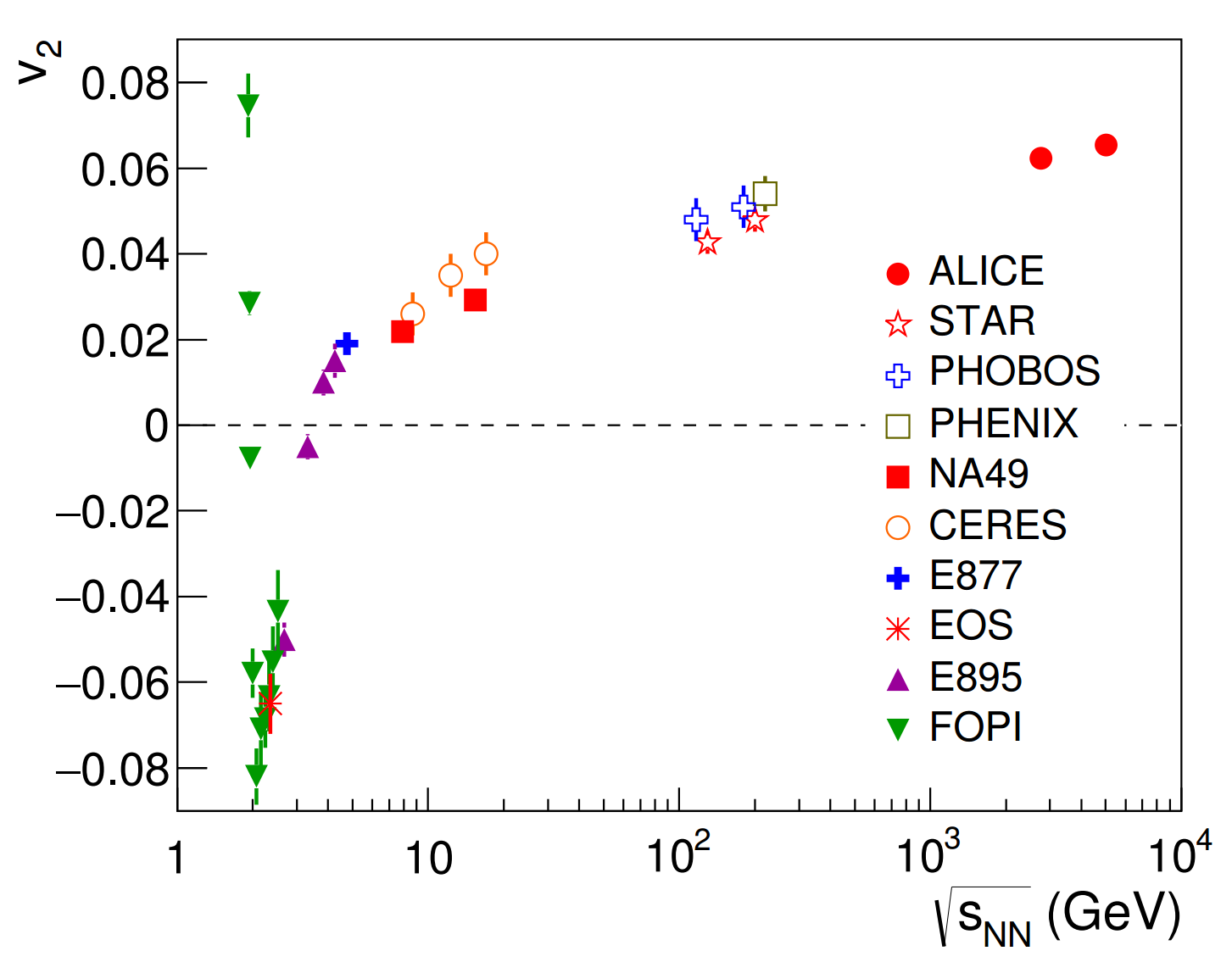}
	\put(-200,130){(b)}
		
	\includegraphics[width=0.81\textwidth,height=4.5cm]{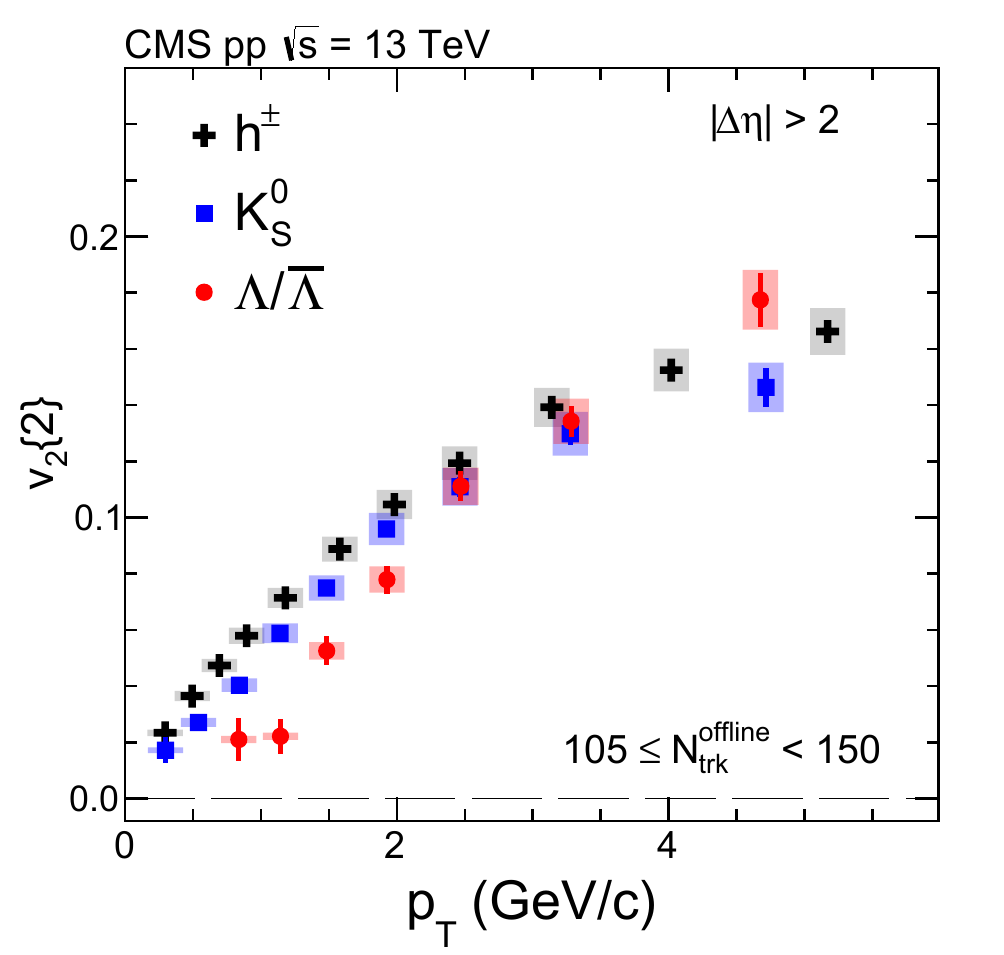}	
	\put(-200,105){(c)}
\end{minipage}
	
	\end{center}
	\caption{
	(a) Top panel: Anisotropic flow $v_{n}$ integrated over the $p_{\rm T}$ range 0.2 $< p_{\rm T} <$ 5.0 GeV/$c$, as a function of event centrality, for two--particle and multi-particle correlations. Measurements for Pb--Pb collisions at \sqrsNN\ 5.02 (2.76) TeV are shown by solid (open) markers ~\cite{ALICE:2011ab}. The ratios of $v_{2}\{2\}$, $v_{2}\{4\}$ and $v_{3}\{2\}$, $v_{4}\{2\}$ between Pb--Pb collisions at 5.02 TeV and 2.76 TeV, are presented in the middle and bottom panels. Hydrodynamic calculations are also presented \cite{Noronha-Hostler:2015uye,Niemi:2015voa}. Figure from~\cite{Adam:2016izf}.	
	(b) Integrated elliptic flow \vtwo\ as a function of collision energy \cite{Voloshin:2008dg,Adam:2016izf}. Figure from~\cite{Adam:2016izf}.
	(c) \vtwo\ coefficient for inclusive charged particles, $K^0_S$ and $\Lambda/\bar{\Lambda}$ particles as a function
of \pT\ in \pp\ collisions at \sqrs 13 TeV  for high-multiplicity events, $105 \leq N^{\rm offline}_{\rm trk} < 150$. Figure from \cite{Khachatryan:2016txc}.
	} 
	\label{fig:flow_ALICE}
\end{figure}
First results on anisotropic flow measurements from \PbPb\ collisions at the highest available energy \mbox{\sqrsNN\ 5.02 TeV}, are shown in Fig.~\ref{fig:flow_ALICE} (a), compared to \PbPb\ collisions at \sqrsNN\ 2.76 TeV.
The centrality dependence of $v_{2}$, $v_{3}$ and $v_{4}$ from two- and multi-particle correlations, 
are shown in the top panel of Fig.~\ref{fig:flow_ALICE} (a).
The \vtwo\ coefficients, $v_{2}\{4\}$, $v_{2}\{6\}$ and $v_{2}\{8\}$ (the number in the curly brackets is the number of particles that are used in correlation), 
estimated from multi--particle correlations, 
less influenced by non-flow effects, 
agree within 1\%, 
which implies that non-flow effects are indeed strongly suppressed. 
The values of $v_{2}\{2\}$ compared to the $v_{2}$ values estimated from  multi--particle  correlations
are higher reflecting the effects of non-flow correlations on $v_{2}\{2\}$.
The $v_{2}$, which originates from the asymmetric pressure gradients of the initial ellipsoidal overlap region,
increases, as expected, with the initial geometric asymmetry from central to peripheral collisions, with maximal value for the centrality range 40--50\%.
The magnitude and centrality dependence of the higher harmonics, $v_3$ and $v_4$, 
reflecting the irregular energy-density fluctuations of the initial stage,  is much weaker.
The values of the measured $v_n$ coefficients are in good agreement with calculations of hydrodynamical models~\cite{Noronha-Hostler:2015uye, Niemi:2015voa}, see solid lines in the top panel of Fig.~\ref{fig:flow_ALICE} (a). 
The moderate increase of $v_{2}$, $v_{3}$ and $v_{4}$, observed between the two energies, 
allows discriminating between different model predictions~\cite{Niemi:2015voa, Noronha-Hostler:2015uye} and parametrizations of the initial conditions 
(in terms of the transport coefficient $\eta/s$, the ratio of the shear viscosity, $\eta$, over the entropy density, $s$).   

\begin{figure}[htbp]
  	\begin{center}
  		\includegraphics[width=\textwidth]{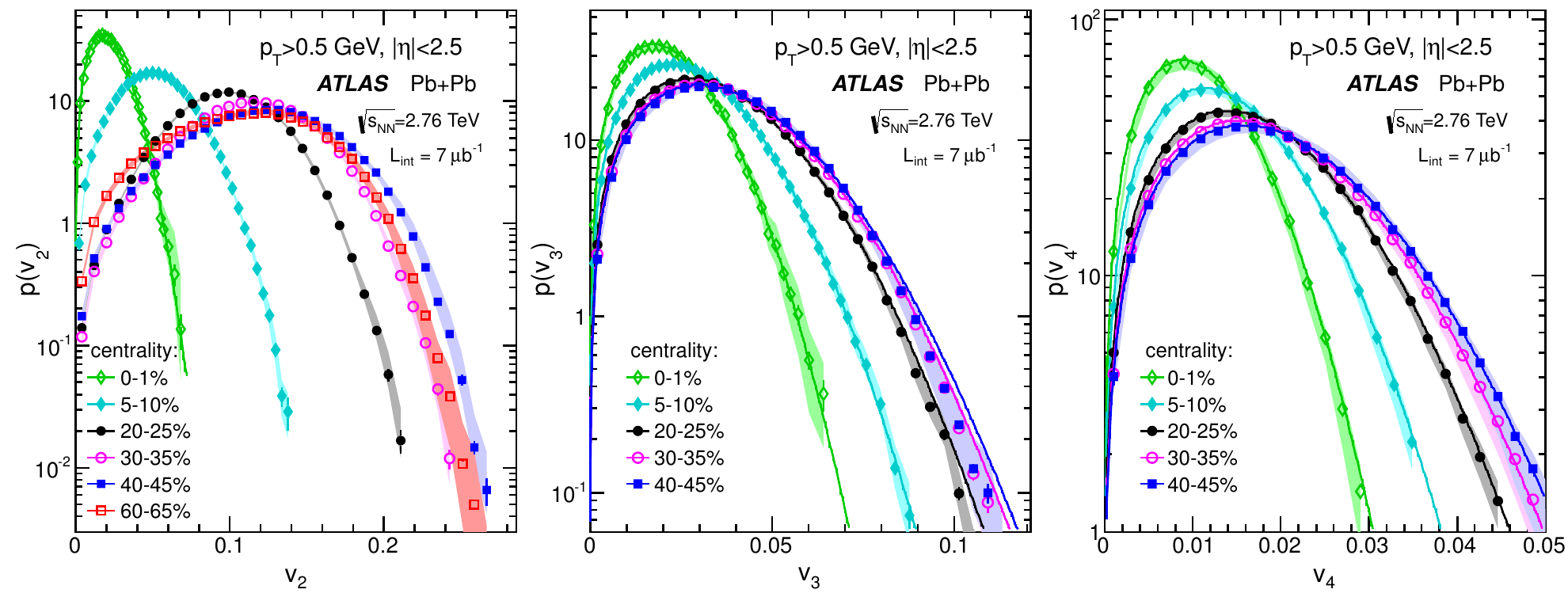}
  		
  	\end{center}
  	
  	\caption{
  	The probability density distributions of the event-by-event anisotropic flow coefficients $v_n$ in several centrality intervals for
  	$n = 2$ (left), $n = 3$ (middle) and $n = 4$ (right). 
  	Solid curves are distributions from fluctuations-only scenario. 
  	Figure from \cite{Aad:2013xma}.} 
  	\label{fig:ATLAS_EbyE_flowVn}
  \end{figure}

Further systematic studies of flow coefficients and their event-by-event fluctuations were performed at LHC~\cite{Aad:2013xma,ALICE:2016kpq}.
Such measurements, are extensively compared to models with the aim to access the fluctuations in the initial state geometry.
The ATLAS results, presented in Fig.~\ref{fig:ATLAS_EbyE_flowVn}, 
show that the $v_{n}$ distributions broaden from central to peripheral collisions, especially for \vtwo, 
reflecting the gradual increase of the magnitude of $v_n$ for more peripheral collisions.  
Moreover, the results are in good agreement with a scenario assuming only fluctuations, except for $v_2$,
confirming that $v_2$ is driven by the initial geometry while the higher harmonics are driven by fluctuations.

Additional constraints on hydrodynamical models, 
and in particular on the initial conditions and transport coefficient
$\eta/s$   
can be set studying the dependence of $v_2$  (and of higher harmonics) on transverse momentum and particle mass.
 Detailed studies of the elliptic flow  $v_2$ of a variety of particles ($\pi$, $K$, $K_{\rm S}^0$, p, $\phi$, $\Lambda$, $\Xi $, $\Omega$) have been carried out~\cite{Abelev:2014pua}. 
 At low \Pt\,  a mass ordering 
related to an interplay of radial and anisotropic flow is seen, that is  
in agreement with hydrodynamical calculations.
Further analysis of identified particles ($\pi$, $K$, p) in \PbPb\ collisions at \sqrsNN\ 5.02 TeV  \cite{Adam:2016nfo} shows that the expected mass ordering is also seen for higher harmonics.

Such precision measurements allow the detailed characterization of the properties of the produced matter.
The parameter that controls most directly the behaviour of the medium,
apart from its equation of state,
is the transport coefficient $\eta/s$ (the relativistic generalisation of the kinematic viscosity).
According to the kinetic theory, 
$\eta$ is proportional to the mean free path of particles in a fluid and unitarity imposes a limit on how small $\eta$ can be at given conditions.
As a consequence of this observation, the quantity $\eta/s$ has a lower bound close to $1/4\pi$, about 0.08 (in units of $\hbar$), obtained in strong-coupling calculations based on the AdS/CFT conjecture~\cite{Kovtun:2004de}. 
The combination of experimental results and their comparisons to hydrodynamical models,
 indicate that the QGP produced at LHC behaves like a strongly-interacting,  
almost perfect liquid with a very low value of the shear viscosity to entropy density ratio ($\eta/s\sim 0.20$),  
close to the theoretical lower  bound.
The current LHC data indicate that $\eta/s$ does not increase significantly in \PbPb\ collisions at \sqrsNN\ 5.02~TeV with respect to \PbPb\ at \sqrsNN\ 2.76~TeV.

The facts that the QGP has been found to behave like an almost ideal fluid with very small viscosity can be considered a gift of nature \cite{Heinz:2013wva}  
because it allows studying the spectrum of the initial-state quantum fluctuations through the experimental measurements of the final-state anisotropic flow fluctuations. In contrast, a high viscosity would imply that all initial-state fluctuations would had been wiped out by dissipation before the final decoupling of the emitted particles. This opens up the possibility to experimentally access the initial state of the collisions and study the quantum nature of the initial energy deposition process.
A comprehensive set of anisotropic flow data were used \cite{Abelev:2012di, Aad:2014fla, CMS:2013bza,CMS:2012xxa}
to compare with calculations 
of different models of the initial energy deposition (which give quite different results; see \cite{Heinz:2013th} for a description of the models and original references) and to over-constraint the dynamical evolution models.
The $v_{\rm n}$ data of CMS \cite{CMS:2013bza} and event-by-event probability distributions measured by ATLAS \cite{Aad:2014fla} cannot be simultaneously described by viscuous hydrodynamics for any choice of $\eta/s$ if initial fluctuation spectra for MC-KLN and MC-Glauber models are used.
More successful in describin the data is the IP-Glasma model \cite{Schenke:2012fw,Schenke:2012wb}.
This model is based on the CGC idea \cite{Kovner:1995ja,Kovchegov:1997ke,Krasnitz:1998ns,Krasnitz:1999wc,Lappi:2003bi,Lappi:2006fp} implementing gluonic field fluctuations inside nucleons \cite{Schenke:2012fw,Schenke:2012wb,Tribedy:2010ab} as well as gluon saturation effects \cite{Kowalski:2003hm,Bartels:2002cj,Kovner:1995ja,Kovchegov:1997ke,Krasnitz:1998ns,Krasnitz:1999wc,Lappi:2003bi,Lappi:2006fp}.
The IP-Glasma initial conditions reproduce the entire measured spectrum of charged hadrons anisotropic flow coefficients $v_{\rm n}$, both integrated over and differential in \pt ,
for all collision centralities as well as the measured \cite{Aad:2013xma} event-by-event distributions of \vtwo , $v_3$ and $v_4$ again for a range of collision centralities \cite{Schenke:2012fw,Schenke:2012wb}
(when evolved with viscous fluid dynamics, after a short initial pre-equilibrium stage modeled by classical Yang-Mills evolution).
In general it is found that the initial fluctuation spectrum can be computed from the CGC theory using the IP-Glasma model and that the gluon field fluctuations inside the nucleons within the colliding nuclei play an essential role in reproducing this spectrum.

The elliptic flow $v_2$ was measured in numerous experiments 
from low to high energies, up to \sqrsNN\ 5.02 TeV.  
The integrated $v_2$, shown in Fig.~\ref{fig:flow_ALICE} (b), 
increases with collision energy, mainly due to the larger mean $p_{\rm T}$~\cite{Aamodt:2010pa}.
In the transition from the highest RHIC to LHC energies, \vtwo\ increases by 30\%~\cite{Aamodt:2010pa}, 
in agreement with hydrodynamic models that include viscous corrections~\cite{Ackermann:2000tr, Luzum:2009sb,Hirano:2010jg,Hirano:2005xf,Drescher:2007uh}. 
However, it should be noted that
various contributions may influence the development of anisotropic flow in heavy-ion collisions,
depending on the different energy domains. 
In general, the hydrodynamic approach was thoroughly tested and emerged as a valid framework over a broad range of collision energies.

Such studies were also extended to the ``small reference systems",  
and important results were obtained on identified particle $v_2$ in \pp\ collisions at \sqrs\
 13 TeV \cite{Khachatryan:2016txc} and \pPb\ collisions at \sqrsNN\ 5.02~TeV \cite{ABELEV:2013wsa,Abelev:2014mda},
that was estimated from the study of two-particle correlations. 
For high-multiplicity \pp\ and \pPb\ collisions a mass ordering, similar to the one seen in \PbPb\ (Fig.~\ref{fig:flow_ALICE} (c)~\cite{Khachatryan:2016txc}), is observed. The striking similarity between \PbPb\ and high-multiplicity \pp\ and \pPb\ results, with the former described by hydrodynamical calculations, may suggest that in the small systems may also have a collective origin at the high LHC energies.

\begin{figure}[htbp]
	\begin{center}
	\includegraphics[width=0.95\textwidth,height=8cm]{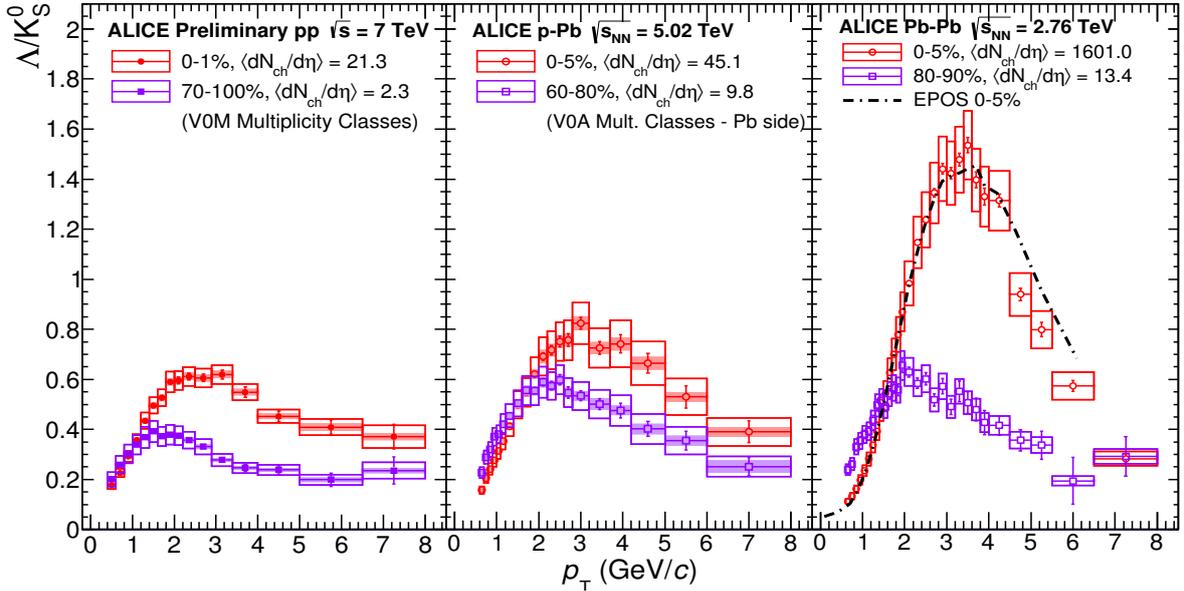}
	\end{center}
	\caption{The $\Lambda/K_{s}^{0}$ ratio for two multiplicity classes in (left) \pp\ collisions at $\sqrt{s}=7$ TeV, (middle) \pPb\ collisions at \sqrsNN\  5.02 TeV and (right) \PbPb\ collisions at \sqrsNN\ 2.76 TeV. The curve shows $\Lambda/K_{s}^{0}$ value for 0--5\% \PbPb\ collisions estimated from EPOS model~\cite{Werner:2012sv}. Figure from \cite{QM2015}.
} 
	\label{fig:lambda_K}
\end{figure}

To further probe collective phenomena, and their possible presence in all measured systems,
detailed studies make use of identified particles and extend to different \Pt\ domains.
In particular, the effects of radial flow on the particle composition at intermediate \Pt\ 
can be investigated studying the baryon-to-meson ratios, $p/\pi$  and
$\Lambda/K_{s}^{0}$ \cite{Abelev:2013xaa,Adams:2006wk,Zhang:2007st}.  
An enhancement of the $\Lambda/K_{s}^{0}$ ratio in AA relative to a similar measurement in pp, 
the so-called ``baryon anomaly", was first observed at RHIC \cite{Adcox:2001mf,Vitev:2001zn}.
The $\Lambda/K_{s}^{0}$ data, measured up
to \pt\ $\approx 8$~GeV/$c$, see Fig.~\ref{fig:lambda_K}, 
confirm that the effect persists at the LHC for \PbPb\ collisions at \sqrsNN\ 2.76~TeV; furthermore, it 
is slightly stronger than at RHIC and extends to higher \Pt.  
This strong rise of the baryon-to-meson ratio at low \Pt\ can be described by relativistic hydrodynamic models
and/or quark recombination from QGP with the  EPOS model~\cite{Werner:2012sv} being particularly  successful in also  consistently describing a multitude of other observables.
A qualitatively similar trend is observed also for \pp\ and \pPb\ collisions, 
as can be seen in the first two panels of Fig.~\ref{fig:lambda_K},
which present  the $\Lambda/K_{s}^{0}$ ratio measured in the most central collisions
compared to the most peripheral centrality range for all measured systems.

Due to space limitations, it is not possible to list here all results of systematic studies in \PbPb , \pPb , and \pp . Overall, all low-\Pt\ observables, studied so far, show flow-like trends with varying magnitudes, which can be interpreted as evidence of collectivity in small systems, and therefore are under intense study, including initial state effects and saturation physics. For more information we point the reader to the recent review~\cite{Loizides:2016tew}. 
We summarize here some of the arguments on the question of possible thermalisation
(which is a prerequisite of collectivity)
in small systems \cite{Schukraft:2013wba}.
Indeed, the question of how such small systems, of a few fm$^3$ compared to about 5000 fm$^3$ in \PbPb , 
could thermalize and develop collective behaviour is being actively investigated.
The proton could be seen as a collection of (sea) partons that can undergo multiparton interactions (MPI),
equivalent to \Ncoll\ collisions of participating nucleons in heavy-ion collisions. 
In addition, the final-state particle densities are large (at least for high-multiplicity events). 
With a possible small thermalisation time of order 1 fm/$c$,
the volume required for thermalisation could be of few fm$^3$.
Once the system equilibrates the lower limit of $\eta/s$ would imply a mean free path close to zero.
Because the applicability of hydrodynamics does not depend on absolute numbers
but it depends on relative sizes, 
the ratio of the system size to the mean free path could be a big number even in small (but dense) systems.   
In such a case viscous hydrodynamics could also be applicable to 
small systems  with conditions providing for a big ratio of the mean free path to the system size.

Overall,  the LHC multi-differential measurements of high-statistics data allowed performing extensive tests of the hydrodynamic framework.
The \PbPb\ results at \sqrsNN\ 2.76 and 5.02 TeV have confirmed its validity and have shown that it is mature enough to make reliable extrapolations and predictions at higher energies.
The hydrodynamic models provided a quantitative characterization of the medium at the two qualitatively similar but distinct \PbPb\ collision energies at LHC, while its applicability in a broader range of colliding systems, including \pPb\ and pp, is being actively investigated.

\vspace{0.5cm}

As a final remark on the studies at the soft sector we mention the 
analysis of dynamical event-by-event fluctuations, proposed as a probe of the QCD phase transition, see the review \cite{Jeon:2003gk}. 
In general, any thermal system displays characteristic thermal fluctuations, and any thermal system with conserved
charges will display characteristic fluctuations in these charges. In hot QCD, many of these fluctuations can be characterized
by susceptibilities that are calculable from first principles on the lattice. Fluctuations are expected to increase if one approaches
a line of first order in the phase diagram, or if one approaches the tricritical point at which this first order line ends (Fig.~\ref{fig:phaseDiagram}-left).
As a consequence, for instance, the beam energy scan at RHIC, that has recently revisited the region of large baryon-chemical
potential in the QCD phase diagram, has focused on these fluctuation measures as a means to search for the existence of
a tricritical point. However, the role of fluctuation measurements is more general. Even in regions of parameter space where
the QCD phase diagram for physical quark masses shows a smooth cross-over, fluctuation measures provide additional
characterizations of the thermal system. For these reasons, dynamical event-by-event fluctuations of the mean \pt\ 
of final-state charged particles  or in net-charge
fluctuations \cite{BraunMunzinger:2011ta},  studied at LHC by ALICE in  \cite{Abelev:2014ckr} and  \cite{Abelev:2012pv} respectively. First results suggest the existence of possible non-trivial collective effects, in line with lattice calculations at high temperatures \cite{Karsch:2001cy}
indicating that the transition from the confined to the deconfined state is a cross-over at the LHC regime.

Further analysis of particle yields at midrapidity, from central \PbPb\ collisions at \mbox{\sqrsNN\ 2.76 TeV}, includes studies of net baryon number and strangeness susceptibilities as well as correlations between electric charge, strangeness and baryon number \cite{Braun-Munzinger:2014lba,Braun-Munzinger:2016miz}.
This work shows that the resulting fluctuations and correlations are consistent with lattice QCD results
at a $T_{\rm c} \sim 155$ MeV which corresponds to the (pseudo)critical temperature  for the chiral
crossover. 
In particular it is worth mentioning that susceptibilities are quantities that can be understood via resumed perturbation theory down to extremely low temperatures, close to $T_{\rm c}$ \cite{Andersen:2012wr,Vuorinen:2002ue}.

The observed agreement of experimental results with Lattice calculations further supports the assumption that the created fireball in such collisions is of thermal origin and exhibits characteristic properties as expected in QCD for the transition from the deconfined plasma of quarks and gluons to the hadronic phase. Because the LHC data show a vanishing baryochemical potential and since the lattice QCD calculations are for  $\mu_b=0$, this is the most direct comparison of the LHC experimental data with theory calculations.

\section{Towards a Heavy-Ion Standard Model}
The new era of precision measurements at LHC,
both in the soft physics sector, discussed in this article,
and for hard observables, discussed in the accompanying article \cite{P2},
is naturally motivating numerous theoretical efforts.
Studying the hot QCD matter at temperatures in the range of up to 2$T_c$ at RHIC 
and in the range of 3$T_c$ at LHC,
is based on a synergy of experiment and theory;
experimental data are largely used to infer the properties of the hot and dense matter
that theory cannot yet reliably predict from first principles QCD.
Experimental measurements pose many theoretical challenges and rise questions stimulating further progress.
The continuous interplay of experiment and theory 
has led to big advances towards developing theoretical frameworks 
where such studies 
can be meaningfully performed 
and a diverse theory ``toolkit" has been developed to target specific questions
aiming to a quantitative characterisation of the produced  matter with controlled uncertainties.
As examples showing the diversity of different approaches and progress on the theoretical front
few are mentioned here.
Semiclassical gauge theory is used to describe the initial conditions reached in nuclear collisions.
Lattice gauge theory is typically employed to study static thermodynamic properties of QCD matter, 
such as its equation-of-state and colour screening.
QCD perturbation theory in the vacuum as well as in a thermal medium is used to describe jets and quarkonia.
Holographic methods, mapping coupled gauge theories on their gravity duals, 
are found to describe the transport properties and the dynamics of thermalisation.
Transport theory and especially viscous hydrodynamics emerged 
as a valid framework for the description of the evolution of the bulk matter \cite{Muller:2013dea}.

Indeed, a ``Standard Model" of the dynamics of ultra-relativistic heavy-ion collisions has emerged in the last decade based on the RHIC results \cite{Muller:2012hr,Heinz:2013wva,Fries:2010ht}.
At LHC multi-differential studies of ``soft observables" 
tested and confirmed the validity of the viscuous hydrodynamics framework
as the ``Standard Model" for 
the description of the dynamical evolution of the QGP medium 
(both at  \sqrsNN\ 2.76 and 5.02 TeV \PbPb\ collisions).
A full framework has been worked out \cite{Muller:2013dea,Gale:2012rq}
including the prediction of the initial conditions for hydrodynamics.

Regarding the initial state, 
the CGC model,
based on the gluon saturation mechanism in the nuclear wave function at
small patron fractional momentum Bjorken-$x$,
can be successfully employed to predict the initial energy and entropy distribution in the nucleus--nucleus collision, and therefore provide the initial conditions for the hydrodynamic evolution.
A complete study of the system evolution  \cite{Gale:2012rq,Gale:2013da} begins with the gluon distribution fluctuations inside the colliding nuclei. Then,  the distributions are evolved  using classical Yang-Mills theory. As a consequence, a fluctuating energy density distribution is obtained and  later injected  into viscous hydrodynamics.
Such an approach provided the means to predict the  behaviour of the produced system for a wide range of collision energies and a wide range of collision systems without arbitrary fit parameters.

One of the conclusions of such a study is that  
the average value of $\eta/s$ (averaged over the thermal history of the expansion)
in \PbPb\ collisions at LHC \sqrsNN\ 2.76 TeV, is 0.20 (with systematic uncertainties of at least 50\%); 
whereas the value for \AuAu\ collisions at the top RHIC energy \sqrsNN\ 200~GeV is 0.12,
suggesting that the average value of $\eta/s$ at the LHC energy (2.76 TeV) 
is approximately 60\% higher than at RHIC \cite{Song:2011qa}
and indicating a temperature dependence of this quantity \cite{Gale:2012rq}.
It also indicates that the QGP at the RHIC lower temperature is more strongly coupled and a better ``perfect" liquid.
First analysis of a limited statistics sample of 
\PbPb\ collisions at \sqrsNN\ 5.02~TeV
indicates that $\eta/s$ does not increase significantly 
with respect to \PbPb\ at \sqrsNN\ 2.76~TeV \cite{Adam:2016izf}.
However, a firm conclusion on the temperature dependence of $\eta/s$,
requires more precise, higher statistics analyses at all studied energies.

On the basis of this  ``Heavy-Ion Standard Model" one can,
in fact, realize the analogies between the evolution of the matter produced in
relativistic heavy-ion collisions, the ``Little Bang" and the matter of the early universe after the ``Big Bang".
In both cases, the initial-state quantum fluctuations propagate to 
macroscopic (measurable) fluctuations in the final state
via the acoustic and hydrodynamic response of the medium.
In the Big Bang expansion, the final-temperature fluctuations probe the bulk dynamics,
photons are the penetrating probes and the light nuclei serve as chemical probes.
In the evolution of the Little Bang, the fluctuations in the final flow profile probe the expansion dynamics,
photons and jets are penetrating probes and the different final-state hadron species serve as chemical probes.

On the other hand, while there is only one ``Big Bang",
systematic studies of the ``Little Bangs," produced at the laboratory in huge numbers,
provide the opportunity to 
test a number of different measurable experimental observables and develop a direct relation to theory,
in particular to QCD calculations on the lattice.  
Such calculations have advanced considerably, 
and progress has also been made even with demanding calculations of dynamic quantities such as transport coefficients.
In more detail, 
(a) the equation-of-sate of the produced medium, which lattice QCD can compute reliably, is reflected in the spectra of the emitted final-state particles.
(b) The transport coefficients of the QGP are related to the final-state flow patterns (discussed in this article) and the energy loss of energetic partons 
(discussed in~\cite{P2}).
In particular, comparisons of experimental data and models made progress in determining the 
(i) the shear viscosity over entropy density ratio, $\eta$/s
(ii) the ``jet quenching parameter",
which is the coefficient governing the transverse momentum diffusion of fast partons for radiational energy loss 
 (iii) the coefficient
 governing collisional energy loss  
 and (iv) the diffusion coefficient of heavy quarks.
 Further progress is needed for lattice gauge theory to reliably calculate all these dynamic quantities.
 (c) The static colour screening length is related to the dissociation of the quarkonia states in the QGP, see~\cite{P2}, 
 which is reliably calculated by lattice QCD.
 (d) The electromagnetic response of the QGP is reflected by the emitted thermal radiation; however, it is challenging to calculate it on the lattice.
In general, it is with a continuous interplay of theory and experiment, that the heavy-ion phenomenology is pursuing the ambitious goal of relating first principle QCD calculations to LHC experimental measurements with a minimal amount of modelling \cite{Muller:2013dea}.

\section{Summary}

We have presented  an overview of results from \PbPb\ collisions at LHC,
as well as from the reference systems \pPb\ and pp, 
focusing on  ``soft observables", 
that are typically used in heavy-ion physics studies
to characterize the bulk matter properties and dynamical evolution of the system created in the collision.

The \PbPb\ results from Run 1 at \sqrsNN\ 2.76 TeV 
and a first glimpse of Run 2 data at \sqrsNN\ 5.02 TeV presented here (and further detailed in recent  conferences i.e.~\cite{SQM2016,ICHEP2016}) show that the QGP medium created in \PbPb\ collisions at LHC is qualitatively similar 
to the one created  in central \AuAu\ collisions at \sqrsNN\ 200 GeV at RHIC.
First results on ``soft observables" from \PbPb\ collisions at \mbox{\sqrsNN\ 5.02 TeV} show that the higher energy of Run 2 
leads to a smooth evolution, in agreement with previously established trends. 
Also according to expectations, the higher collision energies at the LHC  have led to higher energy densities 
(accompanied by higher particle multiplicities)
and consequently to a hotter, bigger and longer-lived medium 
as well as to a more abundant production rate of hard probes.

Hence, the LHC data have provided the opportunity to probe, at higher energies, the ``Heavy-Ion Standard Model", 
as established at RHIC,
test its predictability, and provide experimental input for its possible extensions.
The smooth evolution of many observables from RHIC to LHC energy, 
as well from 2.76 to 5.02 TeV \PbPb\ collisions at LHC,
qualitatively similar but quantitatively different, can still be described by the ``Heavy-Ion Standard Model"
with the ultimate test coming from the measurements of collective flow observables.
The studies of soft observables discussed in this article 
and hard probes discussed in the accompanying article  \cite{P2}
contribute to the detailed characterisation of the QGP at LHC.
The medium created at the LHC is characterized by strong collectivity, and 
it is found to behave, from very early time (on the order of 1 fm/$c$), like an almost perfect liquid. It flows almost unobstructed, reacting to pressure gradients with little internal friction (due to the very small shear viscosity to entropy ratio). It is also extremely opaque to even very energetic coloured particles that propagate through it.

 Overall, such studies, 
 combining simultaneous experimental measurements of soft and hard observables with systematic comparisons to theory,
 provide the means to reveal the nature and properties of the produced QGP matter.
 On one hand, 
 the study of the collective flow properties of the produced matter shows that at thermal momentum scales the QGP is strongly coupled.
 On the other hand, 
 the study of jets in heavy-ion collisions at LHC (and RHIC) implies that at high \Pt\, the QGP is weakly coupled
 (and has a quasiparticle structure).
 It is, therefore, a topic of intense theoretical and experimental research to answer relevant questions; 
 namely, what is the  \Pt\ scale where the transition between strong and weak coupling takes place
 and if the QGP contains quasiparticles at thermal scales.
 
In general,
various theoretical and experimental approaches are being further developed and optimized 
to pursue the remaining open questions.
While relativistic viscous hydrodynamics emerged as a successful description of the evolution of the matter
better experimental and theoretical control over initial conditions is needed to boost further progress.
Developing experimental probes that can determine if the gluons in the initial state 
are weakly or strongly coupled  is essential for further progress in this front.
The current
understanding of the processes transforming the initial quantum state of matter in a relativistic heavy-ion collision into a hydrodynamic fluid has to be improved and better constrained.
While this article focuses on bulk, soft observables, the description of the created state of matter is certainly incomplete without considering the results from hard probes as detailed in the accompanying article  \cite{P2} focusing on results from high-\pt\ and mass measurements.
Such studies developing specific observables and methods 
provide the opportunity 
to fully exploit the new territory that LHC opens up with the highest possible energies and 
are expected to complement the picture provided by the study of soft observables.

In addition, further challenges come from the \pPb\ results, 
which have shown flow-like trends that suggest a possible collective behaviour in pA collisions 
(and which may be present in high-multiplicity \pp\ collisions as well). 
Indeed, the typical hallmarks of collective behaviour in strongly interacting systems 
were recently observed also in high-multiplicity pA  and  \pp\ collisions, 
for all observables that were measured so far and with further measurements being in progress.
Naturally, such experimental results have triggered a lot of interesting developments and debates.
Despite the fact that each individual observation can be reproduced assuming different alternative mechanisms, it is intriguing that the ensemble of observables, typically used to characterize bulk collective properties in AA, shows similar trends for the elementary systems.
Given that all these features had been identified as unique signatures 
of hydrodynamic flow in heavy-ion collisions, the observed similarities between small and large systems, as well as the similarity of the results over a large range of collision energies, 
raises fundamental questions. 
Clearly, before drawing quantitative conclusions, 
further measurements 
need to consider in detail
the complex interplay of hard, semi-hard and soft processes 
and the induced correlations of the studied observables with the  corresponding event activity.
Furthermore, the flow-like results at low \Pt ,
have to be combined with results from hard probes, showing no apparent energy loss at high-\pt, as discussed  in \cite{P2}.
Such considerations are currently the main focus of intense investigations  in \pPb\ collisions pointing to the prospects of creating and studying a droplet of strongly interacting deconfined matter described by hydrodynamics and/or a new state of dense gluon matter at the initial state described by saturation physics. 

In general, understanding these results is expected to have profound implications for the so-called ``heavy-ion" physics, which has been addressing high-temperature QCD  with collisions of heavy-ions so far. At the high collider energies the ``small reference systems" may also provide testing ground of ``heavy-ion" physics.
Such studies are expected to contribute 
to one of the main research objectives of ``heavy-ion" physics; 
namely, to understand how collective phenomena and macroscopic properties,
involving many degrees of freedom, emerge from microscopic laws of elementary-particle physics, 
by applying and extending the Standard Model physics to complex dynamically evolving systems.

In summary, 
the exploration of the phases of strongly-interacting matter under extreme conditions 
is one of the richest topics in the strong-interaction sector of the Standard Model.
Experimental input already brought discoveries and surprises  
and demonstrated the potential of the ``heavy-ion research".
With the increasing statistics that will become available during Run 2 and, with the detector upgrades that will become operational in Run 3 and further in the future, ``heavy-ion" physics will contribute to a more quantitative understanding of the QCD matter under extreme conditions
and continue providing unique and stimulating results.


\section*{Acknowledgements}
We thank Davide Caffarri, Thomas Peitzmann and Urs Wiedemann for useful discussions, and
Roberta Arnaldi, Camelia Mironov, Adam Kisiel, {\L}ukasz Graczykowski, Orlando Villalobos Baillie, Aleksi Vourinen for critical reading of the manuscript and most useful suggestions. 
The work of M. Janik was supported by the Polish National Science Centre under decisions no. 2013/08/M/ST2/00598, no. 2014/13/B/ST2/04054, and no. 2015/19/D/ST2/01600.

\section*{References}
\bibliographystyle{elsarticle-num}
\bibliography{bibliografia}

\end{document}